\documentclass[aps,prd,twocolumn,showpacs,preprintnumbers,nofootinbib,amsmath,amssymb]{revtex4}

\usepackage{graphicx}
\usepackage{dcolumn}
\usepackage{bm}
\usepackage{amsmath}

\newcommand{\beq}{\begin{eqnarray}}
\newcommand{\eeq}{\end{eqnarray}}

\newcommand{\real}{{\sf I}\kern-.12em{\sf R}}
\newcommand{\comp}{{\sf I}\kern-.50em{\sf C}}
\newcommand{\unity}{{\sf I}\kern-.54em{\sf 1}}

\newcommand{\refeq}[1]{Eq.~(\ref{#1})}

\def\spose#1{\hbox to 0pt{#1\hss}}
\def\ltapprox{\mathrel{\spose{\lower 3pt\hbox{$\mathchar"218$}}
 \raise 2.0pt\hbox{$\mathchar"13C$}}}

\begin{document}

\title{Phase Diagram of Yang-Mills Theories in the Presence of a $\theta$ Term}
\author{Massimo D'Elia}
\affiliation{
Dipartimento di Fisica dell'Universit\`a
di Pisa and INFN - Sezione di Pisa,\\ Largo Pontecorvo 3, I-56127 Pisa, Italy}
\email{delia@df.unipi.it}
\author{Francesco Negro}
\affiliation{Dipartimento di Fisica dell'Universit\`a
di Genova and INFN - Sezione di Genova,\\
 Via Dodecaneso 33, I-16146 Genova, Italy}
\email{fnegro@ge.infn.it}
\date{\today}

\begin{abstract}
We study the phase diagram of non-Abelian pure gauge
theories in the presence of a topological $\theta$ term. The dependence 
of the deconfinement temperature on $\theta$ is determined on the 
lattice both by analytic
continuation and by reweighting, obtaining consistent results. 
The general structure of the diagram is discussed on the 
basis of large-$N$ considerations and of the possible 
analogies and dualities existing with the phase diagram of QCD in 
presence of an imaginary baryon chemical potential.
\end{abstract}

\maketitle

\section{Introduction}

The possible presence of a non-zero $\theta$ parameter in the langrangian
of Quantum Chromodynamics (QCD) has been discussed since long.
Such parameter is coupled to the topological charge density,
$$
{\cal L}_\theta = -i\, \theta\, q(x) = -i\, \theta\, \frac{g_0^2}{64\pi^2} \epsilon_{\mu\nu\rho\sigma} F_{\mu\nu}^a(x) 
F_{\rho\sigma}^a(x)\, ,
$$
which violates $P$ and $CP$ symmetries, and
its effects on the structure of non-Abelian gauge theories 
are intimately non-perturbative.

Experimental upper bounds on it are quite stringent,
$|\theta| \lesssim 10^{-10}$.
Nevertheless, the dependence of QCD on $\theta$ is quite
interesting, from both a theoretical and a phenomenological 
point of view; think for instance of the solution to 
the $U(1)_A$ problem, regarding the mass of the $\eta '$ 
meson~\cite{u1wit,u1ven}.

The study of $\theta$ related issues is particularly interesting 
when one investigates the behavior of non-Abelian gauge theories
at finite temperature $T$. Modifications in $\theta$ dependence 
are a probe of the changes in the non-perturbative properties of 
the theory and of the approach to the semiclassical regime expected at 
asymptotically high temperatures~\cite{CDG-78,GPY-81,schafer}.
Topological charge fluctuations may be relevant, also from a phenomenological
point of view, around the deconfinement transition, where local
effective variations of $\theta$ may be detectable as event by event 
$P$ and $CP$ violations in heavy ion collisions~\cite{cme}.

The purpose of the present paper is to discuss some general 
features of the phase diagram of pure $SU(N)$ 
Yang-Mills theories in presence of a $\theta$ term.
Unfortunately, the addition of such a term makes the Euclidean
action complex, hindering direct numerical lattice simulations,
like it happens for QCD at finite baryon chemical potential.
For that reason, most of the present knowledge is based
on model studies,
on the computations of $\theta$ derivatives 
at $\theta = 0$ or on other methods which partially circumvent the 
sign problem, like analytic continuation from imaginary 
chemical potentials~\cite{azcoiti,alles_1,aoki_1,vicari_im,dene,theta_eff}.

In Ref.~\cite{dene} we have already discussed about the dependence of 
the critical deconfining temperature on $\theta$, providing 
an estimate of such dependence in the large $N$ limit and 
a numerical computation for $N = 3$, based on
analytic continuation from results obtained 
at imaginary values of $\theta$, for which 
the action is real.  The main result
is that $T_c$ decreases as a function of $\theta$, being
a linear function of $\theta^2$ for small
$\theta$ values: such 
fact is in agreement with predictions coming 
from continuity based semiclassical approximations~\cite{unsal,poppitz,anber}
and model
computations~\cite{fraga,kouno1,kouno2,kouno3},
and can be simply interpreted by considering that 
the free energy of the confined phase increases,
as a function of $\theta$, more than what happens
for the deconfined phase (since the topological susceptibility 
drops at $T_c$~\cite{susc_ft,lucini_1,vicari_ft}), so that the deconfined phase
becomes more and more favorable as $\theta$ increases~\cite{dene}.

The first purpose of the present study is to provide stronger numerical
evidence regarding the determination of $T_c(\theta)$. Apart
from presenting, in Section~\ref{imaginary}, 
new data at imaginary $\theta$ on a finer lattice,
corresponding to a temporal extent $N_t = 10$, which confirm the
continuum limit extrapolation of Ref.~\cite{dene}, we will
obtain, in Section~\ref{reweighting}, 
an independent determination of 
$T_c(\theta)$ for small values of $\theta$, based on reweighting of 
data at $\theta = 0$, showing that it is  consistent with 
the determination from imaginary $\theta$, hence that systematic 
effects are under control for both methods.
Consistency between analytic continuation and reweighting will be 
demonstrated also for the dependence on $\theta$ of 
other physical observables, like the Polyakov loop.

A question which is naturally related to previous topics is how
physical quantities depend on the topological sector $Q$,
especially around the deconfinement transition. 
Such issue, which is discussed in Section~\ref{fixed},
is of particular interest for the related information about
the systematic effects involved
in numerical simulations carried out within a fixed topological sector,
like those exploiting overlap fermions. The problem has
been investigated by recent literature~\cite{brower,aoki07,cossu},
showing that, for some quantities, systematic effects are well under control,
see for instance Ref.~\cite{cossu} for a study regarding the pure
gauge topological susceptibility at finite $T$. We will show that
for other quantities, like the Polyakov loop, effects on finite volumes 
can be larger,
especially around $T_c$, and that even small deviations of $T_c$
itself are detectable when switching from one sector to the other.

In the last part of the paper, which is contained
in Section~\ref{phdiagram},  
we will discuss the general properties
of the phase diagram in the $T_c - \theta$ plane.
Unfortunately, presently available numerical methods, like analytic 
continuation or reweighting, do not permit to obtain much
reliable information, apart from the curvature of the 
critical line $T_c(\theta)$ at $\theta = 0$. Therefore, part 
of the discussion
is based on known large-$N$
considerations and model predictions~\cite{wit80,ohta,wit98,ztn1,wit982,saksug,bergman,ztn2,ztn2b,ztn3}, as well as  
recent numerical evidence~\cite{BDPV} regarding 
the change in the realization of $\theta$ dependence and
periodicity, which takes
place at the deconfinement transition.
Particular emphasis will be placed on the analogy 
that we draw between the
$T_c - \theta$ diagram and the phase diagram 
of QCD in presence of an imaginary baryon chemical potential
$\mu_B$: we will speculate about the duality between the two diagrams,
in the sense of an exchange between the high-$T$ and the low-$T$ regions,
and about its possible relation 
with a duality of the relevant degrees of freedom.
 Finally, in Section~\ref{conclusions}, we will draw our conclusions.

\section{Dependence of the deconfining temperature on $\theta$}
\label{tctheta}

Various model computations predict that the critical deconfining 
temperature in QCD decreases as the $\theta$ parameter 
is switched on~\cite{unsal,dene,poppitz,anber,kouno1,kouno2,kouno3}.
The theory is CP-invariant for $\theta = 0$, hence
thermodynamical quantities and $T_c(\theta)$ itself are 
expected to be even functions of $\theta$, 
i.e. 
\beq
\frac{T_c(\theta)}{T_c(0)} = 1 - R_\theta\ \theta^2 + O(\theta^4)
\label{Tcdep}
\eeq
if the theory is analytic around
$\theta = 0$.

A decreasing $T_c(\theta)$ means that the curvature
$R_\theta$ is positive. 
A possible argument to understand such decrease has been 
given in Ref.~\cite{dene}.
The free energy increases as a function of $\theta$, 
and the coefficient of the lowest order term, which
is quadratic in $\theta$, is given by $\chi/2$, where
$\chi$ is 
the topological susceptibility
($\chi \equiv \langle Q^2 \rangle/(a^4 V)$ and $a^4 V$ 
is the space-time volume).
Due to the sharp drop of 
$\chi$ across the deconfinement transition~\cite{susc_ft,lucini_1,vicari_ft}, 
the increase of free energy
in the confined phase is larger than that in the deconfined phase;
hence, as $\theta$ increases, 
it becomes more and more favorable to the system 
to stay in the deconfined phase, so that 
the deconfining temperature moves to lower temperatures. In particular, 
for a first order transition, which 
is the case for $SU(N)$ pure gauge theories with 
$N \geq 3$, one finds~\cite{dene}:
\beq
\frac{T_c (\theta)}{T_c (0)} = 
1 - \frac{\Delta \chi}{2 \Delta \epsilon} \theta^2 + O(\theta^4)  
\label{prediction}
\eeq  
where $\Delta \epsilon$ and $\Delta \chi$ are respectively
the jump of the energy density and the drop of the
topological susceptibility at the transition.
In the large $N$ limit, 
$\Delta \chi$ tends to $\chi$ computed at $T = 0$ and 
stays finite, while
$\Delta \epsilon \propto N^2$, so that $R_\theta \propto 1/N^2$.

The first numerical results regarding $R_\theta$ have been given 
in Ref.~\cite{dene} for the $SU(3)$ pure gauge theory,
exploiting the idea of performing simulations at imaginary
values of $\theta$ in order to avoid 
the sign problem~\cite{azcoiti,alles_1,aoki_1,vicari_im}.
The approach is the same adopted for QCD at finite baryon chemical 
potential $\mu_B$, 
where purely imaginary values of $\mu_B$ avoid complex values
of the fermion determinant: one can then make 
use of 
analytic continuation to infer the dependence at
real $\mu_B$, at least for small values of $\mu_B/T$~\cite{immu};
in particular the critical temperature
can be reliably estimated up to the quadratic order in 
$\mu_B$, while ambiguities related to the procedure of analytic
continuation may affect higher order terms~\cite{immu_cea}.
The same approach can be used to explore physics at
non-zero $\theta$, if one 
assumes that the theory is analytic around $\theta = 0$,
a fact supported by our present knowledge about 
free energy derivatives at 
$\theta = 0$~\cite{vicari_b4,nostro_b4,alles_2,giusti,vicari_rep,BDPV}.

In Ref.~\cite{dene}, the curvature $R_\theta$ has been determined
by analytic continuation on three different lattice sizes,
$16^3 \times 4$, $24^3 \times 6$ and $32^3 \times 8$, corresponding
to the same physical spatial volume and different 
lattice spacings, $a \simeq (4T_c)^{-1}, (6T_c)^{-1}$ and $(8T_c)^{-1}$,
around the transition. That has permitted us to extrapolate 
the curvature to the continuum limit, obtaining $R_\theta = 0.0175(7)$,
a value which is in rough agreement with the model prediction
in Eq.~(\ref{prediction})~\cite{dene}.

In the present study we make progress by performing new numerical
simulations, both at zero and non-zero imaginary $\theta$, on 
a $40^3 \times 10$ lattice. On the one hand, in Section~\ref{imaginary}
we obtain
a new determination of $R_\theta$ by analytic continuation, on a finer
lattice, which permits us to check and improve the continuum extrapolation
of Ref.~\cite{dene}. On the other hand, the determination of the 
topological background $Q$ of configurations sampled at $\theta = 0$
will permit us to obtain direct information at real $\theta$ 
by reweighting techniques, as illustrated in Section~\ref{reweighting}:
in this way we shall be able to check the reliability of analytic continuation
and to put the numerical determination of $R_\theta$ on a more solid basis.
Finally Section~\ref{fixed} is devoted to investigate the dependence
of physical quantities, including the critical temperature,
on the topological sector.

\subsection{Results from imaginary $\theta$}

\label{imaginary}

The partition function of lattice $SU(N)$ gauge theories 
in presence of an imaginary theta term reads
\begin{equation}
Z_L(T,\theta_L) = 
\int [dU]\ e^{ -S_L [U] - \theta_L Q_L[U]} \, ,
\label{partfunlat}
\end{equation}
where $U$ stands for a configuration of gauge link variables,
$U_\mu (n)$, while $S_L$ and $Q_L$ are the 
lattice discretizations of respectively the 
pure gauge action and the topological charge, 
$Q_L = \sum_x q_L(x)$. As in Ref.~\cite{dene}, we consider the Wilson
plaquette action and a simple
dicretization for $q_L$:
\beq
q_L(x) = {{-1} \over {2^9 \pi^2}} 
\sum_{\mu\nu\rho\sigma = \pm 1}^{\pm 4} 
{\tilde{\epsilon}}_{\mu\nu\rho\sigma} \hbox{Tr} \left( 
\Pi_{\mu\nu}(x) \Pi_{\rho\sigma}(x) \right) \; ,
\label{eq:qlattice}
\eeq
where $\Pi_{\mu\nu}$ is the plaquette operator,
${\tilde{\epsilon}}_{\mu\nu\rho\sigma} = {{\epsilon}}_{\mu\nu\rho\sigma}
$ for positive directions  and ${\tilde{\epsilon}}_{\mu\nu\rho\sigma} =
- {\tilde{\epsilon}}_{(-\mu)\nu\rho\sigma}$. With this choice, gauge links
still appear linearly in the modified action, hence a standard heat-bath 
algorithm over $SU(2)$ subgroups, combined with over-relaxation, can be 
implemented; that would not be possible for different improved 
choices of $q_L$,
like for instance smeared or fermionic operators.

In general, the lattice operator $q_L(x)$ is linked
to the continuum $q(x)$ 
by a finite multiplicative renormalization~\cite{zetaref}
\beq
 q_L(x) {\buildrel {a \rightarrow 0} \over \sim} a^4 Z(\beta) q(x) + O(a^6) \; ,
\label{eq}
\eeq
where $a = a(\beta)$ is the lattice spacing and $\lim_{a \to 0} Z = 1$. 
Hence, as the continuum limit is approached, 
the imaginary part of $\theta$ is related to the lattice parameter
$\theta_L$ appearing in Eq.~(\ref{partfunlat}) 
as follows:
$\theta_I = Z\, \theta_L$.

Knowledge about $Z(\beta)$ is essential to fix the working physical value
of $\theta_I$. It is important to stress that other renormalizations,
linked to the choice of the lattice operator $q_L$, may affect the
free energy, e.g. in the form of additive renormalizations stemming
from two (or more) point correlators of $q_L$. Such UV terms, 
however, are continuous across the phase transition, hence they do
not play any role in the determination of $T_c$ as a function of 
$\theta_I$: this is confirmed by the fact that, as shown in Ref.~\cite{dene}
and in the present manuscript, 
a consistent extrapolation to the continuum can be taken for
$R_\theta$.

Here we will make use
of the non-perturbative determination of $Z$ reported in Ref.~\cite{dene}, see in particular
Fig.~2 reported therein. That has been obtained by measuring, on symmetric
$T = 0 $ lattices, the following quantity~\cite{vicari_im}
\beq
Z \equiv \frac{\langle Q Q_L \rangle}{\langle Q^2 \rangle}
\eeq
where $Q$ is, configuration by configuration, the integer closest to
the topological charge obtained after cooling~\cite{vicari_b4,cooling}.
The idea is similar to that used by heating techniques~\cite{ref:heating}, 
where the average value of $Q_L$ is determined within a fixed topological sector.

\begin{figure}[t!]
\vspace{0.4cm}
\includegraphics*[width=0.47\textwidth]{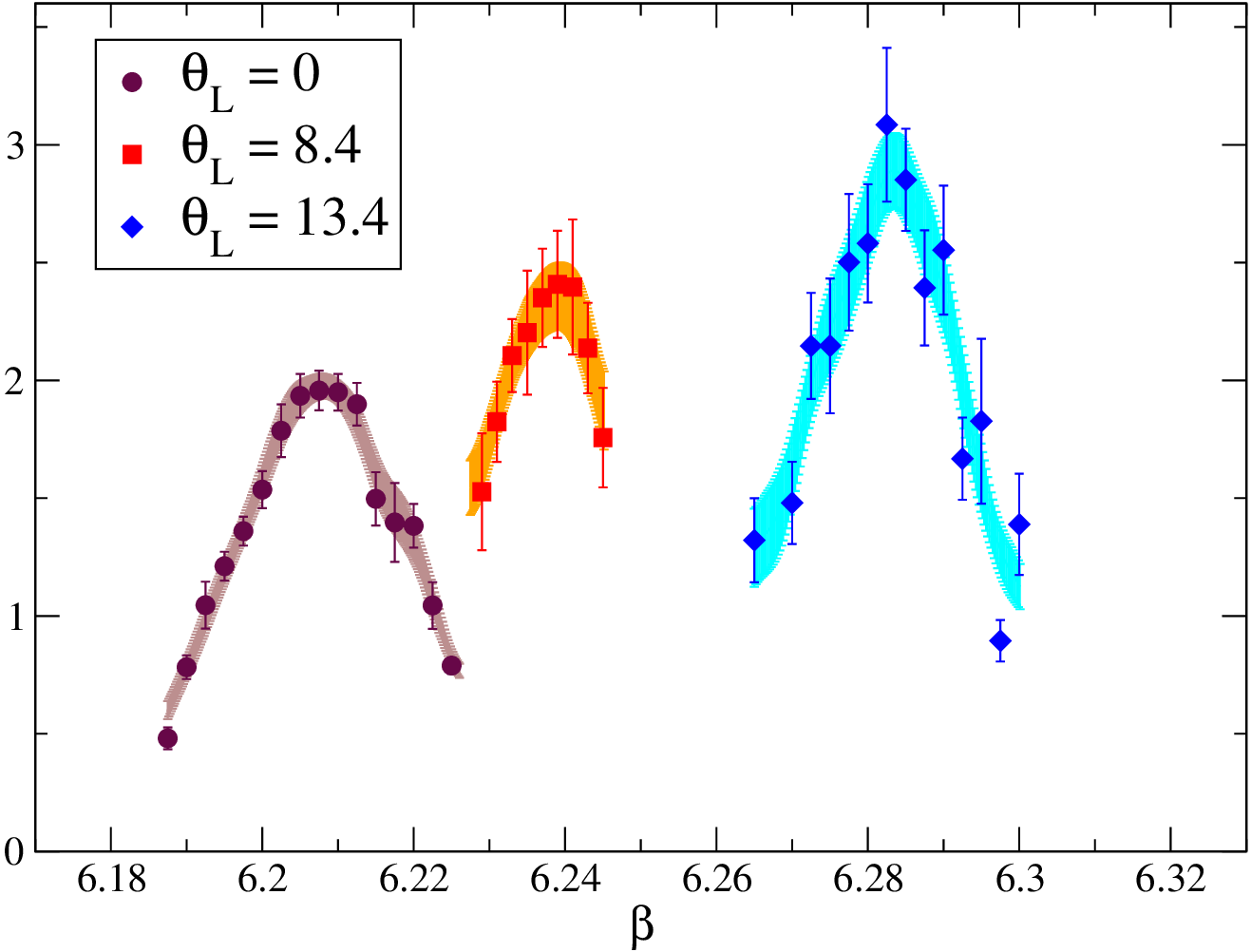}
\caption{Polyakov loop susceptibility as a function of 
$\beta$ on the $40^3 \times 10$ lattice for some explored values of
$\theta_L$.}
\label{fig:pol}
\end{figure}

\begin{table}
\begin{center}
\begin{tabular}{|c|c|c|c|c|}
\hline
lattice & $\theta_L$ & $\beta_c$ & $\theta_I$ & $T_c(\theta_I)/T_c(0)$ \\ \hline
$16^3 \times 4\ $  &   0\ & 5.6911(4)\ & 0\ & 1 \\
\hline
$16^3 \times 4\ $  &   5\ & 5.6934(6)\ & 0.370(10)\ & 1.0049(11)\ \\
\hline
$16^3 \times 4\ $  &  10\ & 5.6990(7)\ & 0.747(15)\ & 1.0171(12)\ \\
\hline
$16^3 \times 4\ $  &  15\ & 5.7092(7)\ & 1.141(20)\ & 1.0395(11)\ \\
\hline
$16^3 \times 4\ $  &  20\ & 5.7248(6)\ & 1.566(30)\ & 1.0746(10)\ \\
\hline
$16^3 \times 4\ $  &  25\ & 5.7447(7)\ & 2.035(30)\ & 1.1209(10)\ \\
\hline
\hline
$24^3 \times 6\ $  &  0\   & 5.8929(8)\   & 0\ & 1\ \\
\hline
$24^3 \times 6\ $  &  5\   & 5.8985(10)\ & 0.5705(60)\ & 1.0105(24)\ \\
\hline
$24^3 \times 6\ $  &  10\ & 5.9105(5)\  & 1.168(12)\ & 1.0335(18)\ \\
\hline
$24^3 \times 6\ $  &  15\ & 5.9364(8)\   & 1.836(18)\ & 1.0834(23)\ \\
\hline
$24^3 \times 6\ $  &  20\ & 5.9717(8)\   &  2.600(24)\ & 1.1534(24)\ \\
\hline
\hline
$32^3 \times 8\ $  &   0\ & 6.0622(6)\   & 0\ & 1\  \\
\hline
$32^3 \times 8\ $  &   5\ & 6.0684(3)\   & 0.753(8)\ &  1.0100(11)\     \\
\hline
$32^3 \times 8\ $  &   8\ & 6.0813(6)\   & 1.224(15)\ &  1.0312(14)\     \\
\hline
$32^3 \times 8\ $  &   10\ & 6.0935(11)\ & 1.551(20)\ &  1.0515(21)\  \\
\hline
$32^3 \times 8\ $  &   12\ & 6.1059(21)\ & 1.890(24)\ & 1.0719(34) \\
\hline
$32^3 \times 8\ $  &   15\ & 6.1332(7)\   & 2.437(30)\ & 1.1201(17)\ \\
\hline
\hline
$40^3 \times 10\ $  &   0\ & 6.2082(4)\   & 0\ & 1\  \\
\hline
$40^3 \times 10\ $  &   6\ & 6.2236(8)\   & 1.068(7)\ &  1.0232(14)\     \\
\hline
$40^3 \times 10\ $  &   8.4\ & 6.2381(5)\   & 1.509(10)\ &  1.0453(10)\     \\
\hline
$40^3 \times 10\ $  &   13.4\ & 6.2821(9)\   & 2.461(22)\ &  1.1144(16)\     \\
\hline
\end{tabular}
\end{center}
\caption{Collection of results obtained for $\beta_c$ and $T_c$. Results for $N_t = 4,6,8$ are taken from Ref.~\cite{dene} and reported for completeness.}
\label{tab:res}
\end{table}

\begin{figure}[t!]
\vspace{0.4cm}
\includegraphics*[width=0.47\textwidth]{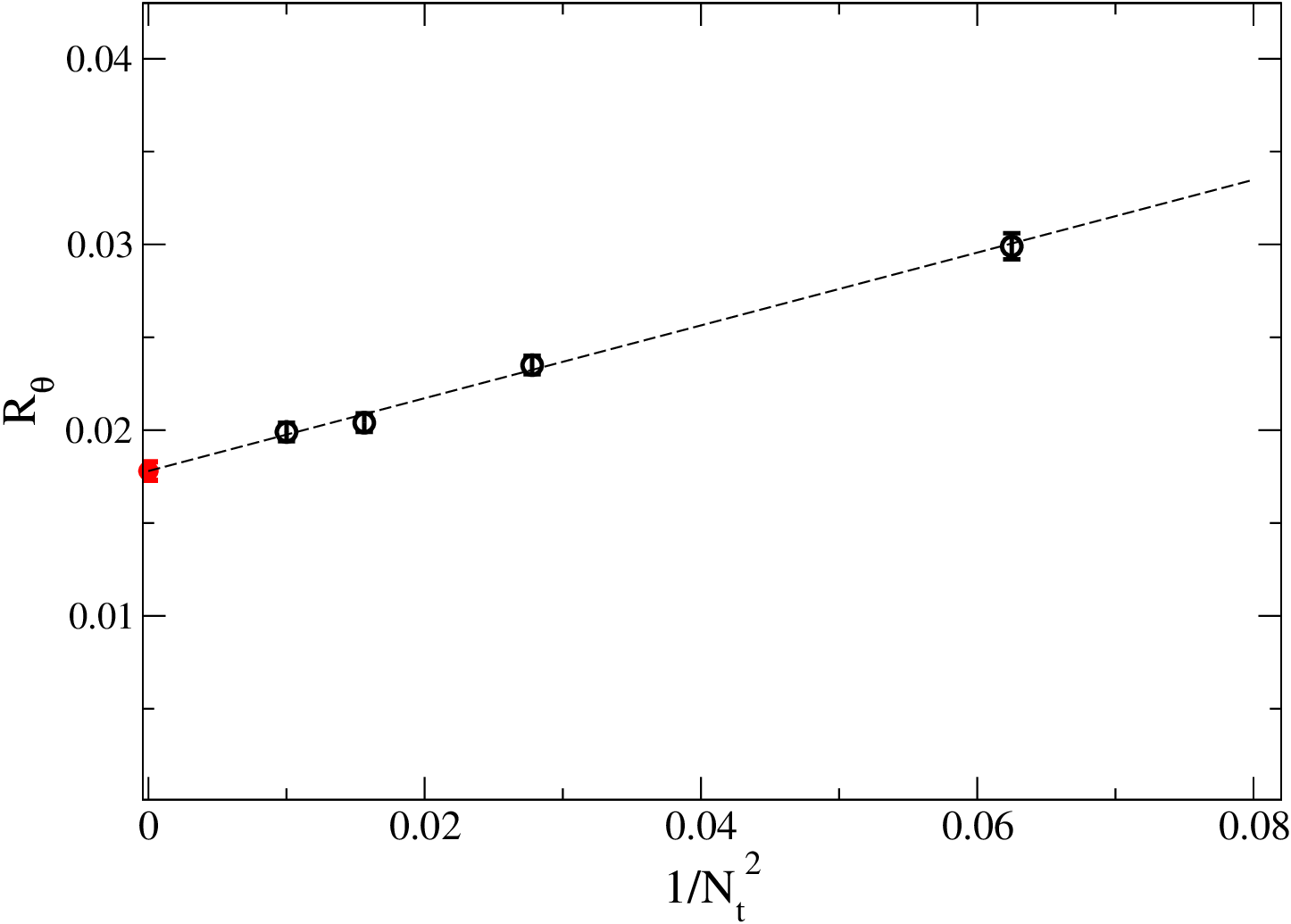}
\caption{$R_\theta$ as a function of $1/N_t^2$. The point at $1/N_t = 0$ is the continuum limit
extrapolation, assuming
$O(a^2)$ corrections.}
\label{fig:contlim}
\end{figure}

The new set of numerical simulations on the $40^3 \times 10$ lattice 
have been carried out at four different values of $\theta_L$,
$\theta_L=0.0,\, 6.0,\, 8.4$ and $13.4$.
We have performed several series of simulations
at fixed $\theta_L$ and variable $\beta$. 
Typical statistics have been of $O(10^5)$ measurements 
per $\beta$ at $\theta =0$ and of $O(10^4)$ measurements 
per $\beta$ at $\theta \neq 0$,
each separated by an updating cycle of 
4 over-relaxation + 1 heat-bath sweeps. The somewhat larger statistics at 
$\theta = 0$ is justified in view of the further analysis reported
in the following subsections.
The numerical effort required for this new $N_t = 10$ lattice is 
significantly larger than that made in Ref.~\cite{dene}, both 
because of the larger lattice size and because of the 
larger autocorrelation times, going up to 
a few hundred cycles around the transition.  

As in Ref.~\cite{dene}, 
in order to determine the deconfinement transition temperature,
we have considered the Polyakov loop and its susceptibility
\beq
L 
&\equiv& \frac{1}{V_s} \sum_{\vec{x}} \frac{1}{N} {\rm Tr}\
       \prod_{t=1}^{N_t} U_0({\vec x},t)  \, \nonumber \\
\chi_L &\equiv& V_s\ 
(\langle |L|^2 \rangle - \langle |L| \rangle^2 )
\rangle \, , 
\label{obs}
\eeq
where $V_s$ is the spatial volume and $|L|$
is the Polyakov loop modulus.
Center symmetry, which corresponds
to a multiplication of all
parallel transports at a fixed time by an element of the center
of the $SU(N)$ gauge group, $Z_N$,
is spontaneously broken at the deconfinement transition and
the Polyakov loop, which is not invariant under 
center transformations, is a related order parameter.
The modified action $S_L + \theta_L Q_L$ is also center symmetric, hence
the Polyakov is an exact order parameter also at $\theta \neq 0$.

The 
Polyakov loop susceptibility is plotted as a function
of $\beta$ in Fig.~\ref{fig:pol},
 together with data obtained after reweighting in $\beta$.
As $\theta_L$ increases, the susceptibility peak moves to 
higher values of $\beta$, i.e. to higher 
temperatures $T = 1/(a(\beta) N_t)$.

The critical couplings $\beta_c(\theta_L)$ have been obtained by performing a 
Lorentzian fit to the unreweighted data of the susceptibility. 
From $\beta_c(\theta_L)$ we reconstruct
$T_c(\theta_L)/T_c(0) = a(\beta_c(0))/a(\beta_c(\theta_L))$ by means of 
the non-perturbative determination of $a(\beta)$ reported in 
Ref.~\cite{karsch_thermo}; in general the location of 
$T_c$ is affected by finite size corrections, which however 
should almost cancel when computing the ratio $T_c(\theta_L)/T_c(0)$.
Finally, $\theta_L$ must be converted into $\theta_I$ by exploiting
the determination of $Z$ at the critical coupling $\beta_c$,
which is obtained by interpolation of data reported in Ref.~\cite{dene}.
All results are shown in Table~\ref{tab:res}, where we also report, for the 
reader's convenience, data obtained
on different lattices in Ref.~\cite{dene}.

The values obtained for $T_c(\theta_I)/T_c(0)$ can be fitted according
to Eq.~(\ref{Tcdep}), with $\theta_I^2 = - \theta^2$. 
If we restrict to $\theta_I < 2$, we get  $R_\theta= 0.0200(5)$,
with $\chi^2/{\rm d.o.f.} \simeq 0.11 $. 
As a final step, we can put the value of $R_\theta$ together
with those obtained for smaller values of $N_t$ in Ref.~\cite{dene},
see Fig.~\ref{fig:contlim}. An extrapolation to the continuum limit
assuming $\mathcal{O}(a^2)$ corrections, 
$R_\theta(N_t) = R_{\theta}^{\textnormal{cont}}+b/N_t^2$,
yields 
$R_{\theta}^{\textnormal{cont}} = 0.0178(5) $, with $\chi^2/{\rm d.o.f.} 
\simeq 0.6 $. That is consistent with 
the continuum extrapolation reported in 
Ref.~\cite{dene}, and in rough agreement with 
the leading $1/N$ estimate for $SU(3)$,
$R_\theta = 0.028(6)$~\cite{dene}. 
The most significant correction to
the large-$N$ prediction can be attributed to the fact that,
for finite $N$, the susceptibility does not drop sharply to zero
at the transition, i.e. $\Delta \chi$ in Eq.~(\ref{prediction})
is less than the value of $\chi$ in the confined phase.

\begin{figure}[t!]
\vspace{0.46cm}
\includegraphics*[width=0.47\textwidth]{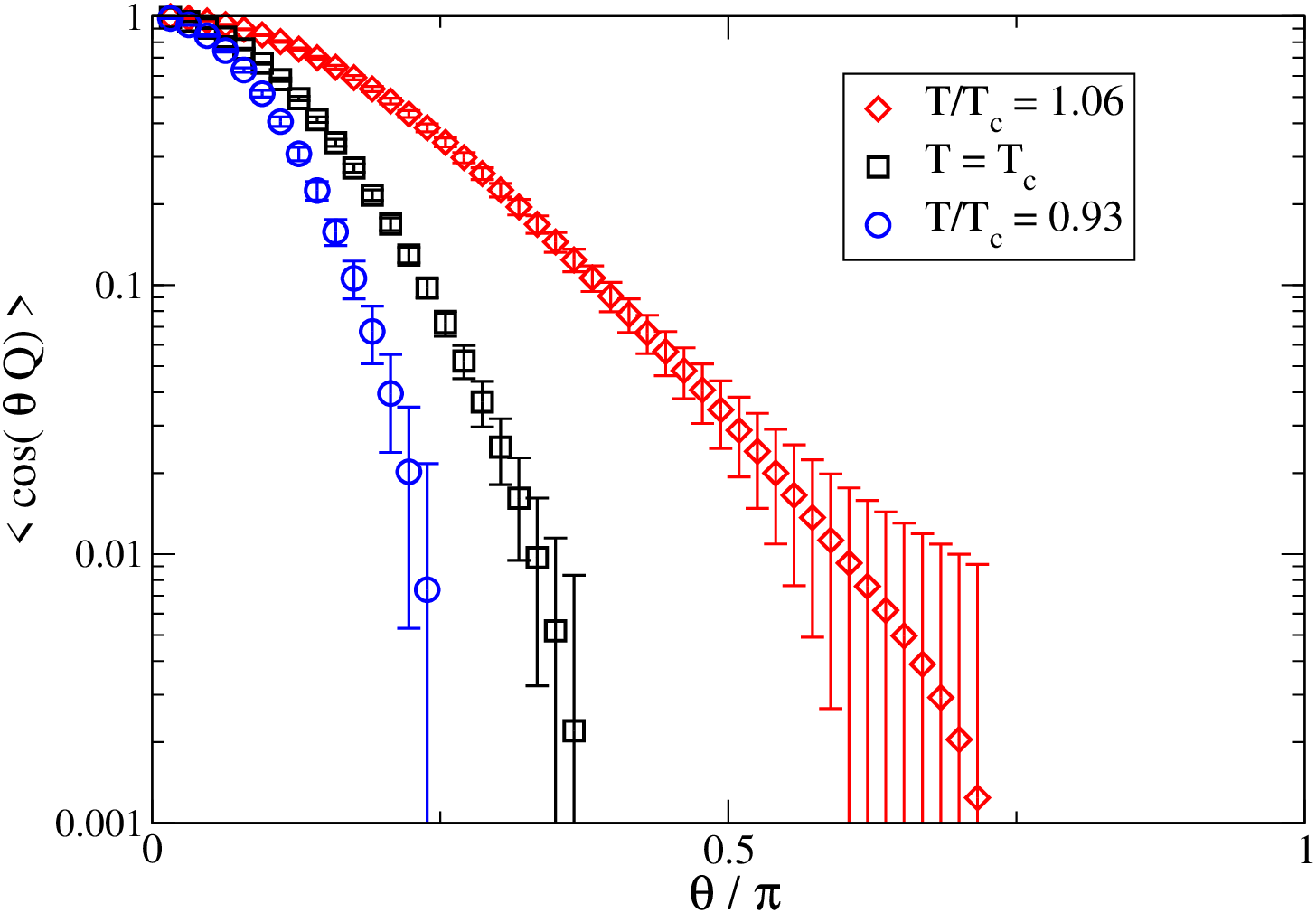}
\caption{Dependence of  $\langle\cos(\theta Q)\rangle$ on $\theta$ on the $40^3 \times 10$ lattice and for three different values of $T$.}
\label{fig:coseno}
\end{figure}

\subsection{Comparison with reweighting at real $\theta$}
\label{reweighting}

Presently known solutions to the sign problem are only
approximate and tipically introduce assumptions and
systematic errors. In the case of analytic continuation
an obvious assumption is that of analyticity
around $\theta = 0$. 
A possible way to keep such effects under
control is to compare different, independent methods,
cross-checking results. 

A method alternative to analytic continuation, which has been 
largely used in QCD at finite baryon chemical potential, is reweighting.
The idea is to sample configurations at $\theta = 0$ and
to move the complex factor of the path integral measure
into the observable 
, i.e., for a generic quantity $O$,
\begin{equation}
\langle O \rangle_\theta = \frac{\int [dU]\ e^{ -S_L [U] + i \theta Q}\,  O }
{\int [dU]\ e^{ -S_L [U] + i \theta Q}} 
= \frac{\langle e^{i \theta Q} O \rangle_{}  }
{\langle \cos(\theta Q) \rangle_{}}
\, .
\label{eq:rewgen}
\end{equation}
where averages without subscript are taken as usual at $\theta = 0$, and
the equality
$\langle e^{i \theta Q} \rangle_{} 
= \langle \cos(\theta Q) \rangle_{}$
has been used,
which derives from the symmetry under $Q \to -Q$ of the 
distribution at $\theta = 0$. 
The major drawback of reweighting is that configurations
sampled at $\theta = 0$ may not be representative enough of the physics
at $\theta \neq 0$; such problem gets worse and worse as
$\theta$ increases and as the thermodynamical limit is approached.
A measure of the severeness of the problem is given 
by the average phase factor in 
the denominator of Eq.~(\ref{eq:rewgen}): 
as $\langle \cos(\theta Q) \rangle$ 
vanishes, one would need unfeasibly large statistics to keep
statistical errors under control. Such problems are well known
from QCD at finite baryon density~\cite{barbour};
a partial improvement can be achieved by reweighting 
in more than one parameter~\cite{fodor}.

Since in the reweighting method the topological charge does
not enter the sampling algorithm directly, one can make use 
of  smoothed gluonic or fermionic 
definitions of $Q$, in order to 
avoid issues related to renormalization. However, the implementation
must be cheap enough to permit the collection
of a sufficiently large sample of measures. 
We have adopted cooling, in particular the implementation outlined 
in Ref.~\cite{vicari_b4}, which is known to provide 
reliable results on fine enough lattices.
This is the reason why we have decided to apply 
the reweighting method only to 
configurations sampled on the $N_t = 10$ lattice.
$Q$ has been measured once every 4 updating cycles;
we will show results for $Q$
obtained after  $n_{cool} = 30$ cooling sweeps, however we have
checked that different choices 
lead to compatible
results.

\begin{figure}[t!]
\vspace{0.46cm}
\includegraphics*[width=0.47\textwidth]{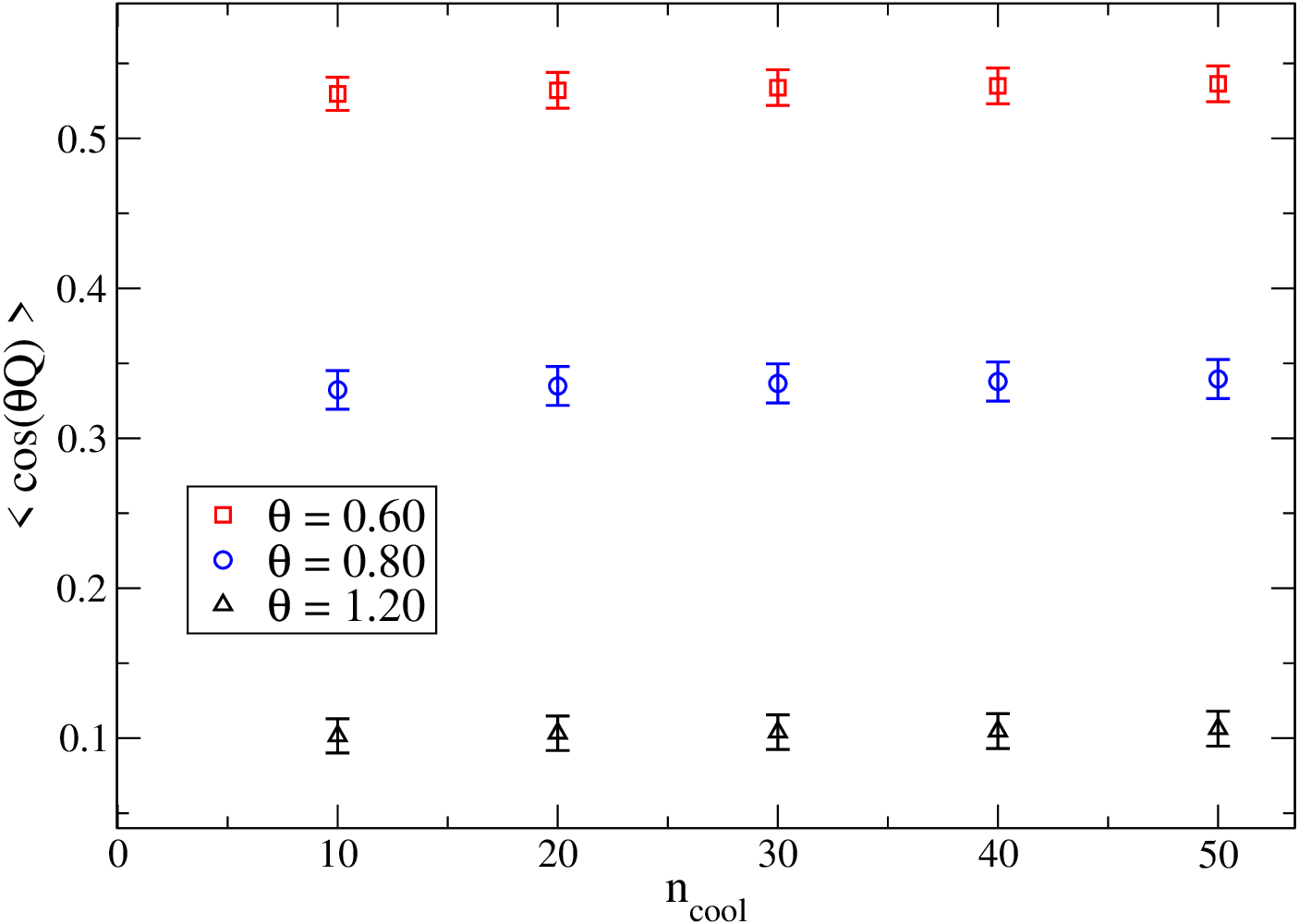}
\caption{Dependence of  $\langle\cos(\theta Q)\rangle$ on the number
of cooling steps 
for $T \simeq 1.06\, T_c$ and three values of $\theta$.}
\label{fig:coscool}
\end{figure}

Let us start the discussion of our results by showing, 
in Fig.~\ref{fig:coseno}, the behavior of the average
phase factor, $\langle\cos(\theta Q)\rangle_{}$,
as a function of $\theta$,  
for 3 different bare couplings, $\beta = 6.1600,\,6.2075,$ and $6.2475$, 
corresponding to $T \simeq
0.93 \,T_c $, $\,T_c $ and $1.06 \,T_c $ respectively.
In Fig.~\ref{fig:coscool} we also 
show $\langle\cos(\theta Q)\rangle_{}$ as a function 
of $n_{cool}$ for a few values of $\theta$ 
at $T \simeq 1.06\, T_c$, which nicely demonstrates 
the stability of results under different choices of $n_{cool}$.

The regions where $\langle\cos(\theta Q)\rangle_{}$ becomes very 
small are hardly accessible to reweighting.
It is clear that the situation is worse in the confined phase, where 
only $\theta \lesssim 0.2\, \pi$ seems accessible, than in the deconfined phase,
where $\theta \sim 0.5\, \pi$ seems reachable. That can be 
understood in terms of the much lower topological activity present
in the deconfined phase. It should be stressed, however, that as the 
thermodynamical limit, $V_s \to \infty$, is taken, arbitrarily large
fluctuations of the global charge $Q$ are expected in both phases,
so that $\langle\cos(\theta Q)\rangle_{}$ must drop to 
zero for any $\theta \neq 0$.

\begin{figure}[t!]
\vspace{0.3cm}
\includegraphics*[width=0.47\textwidth]{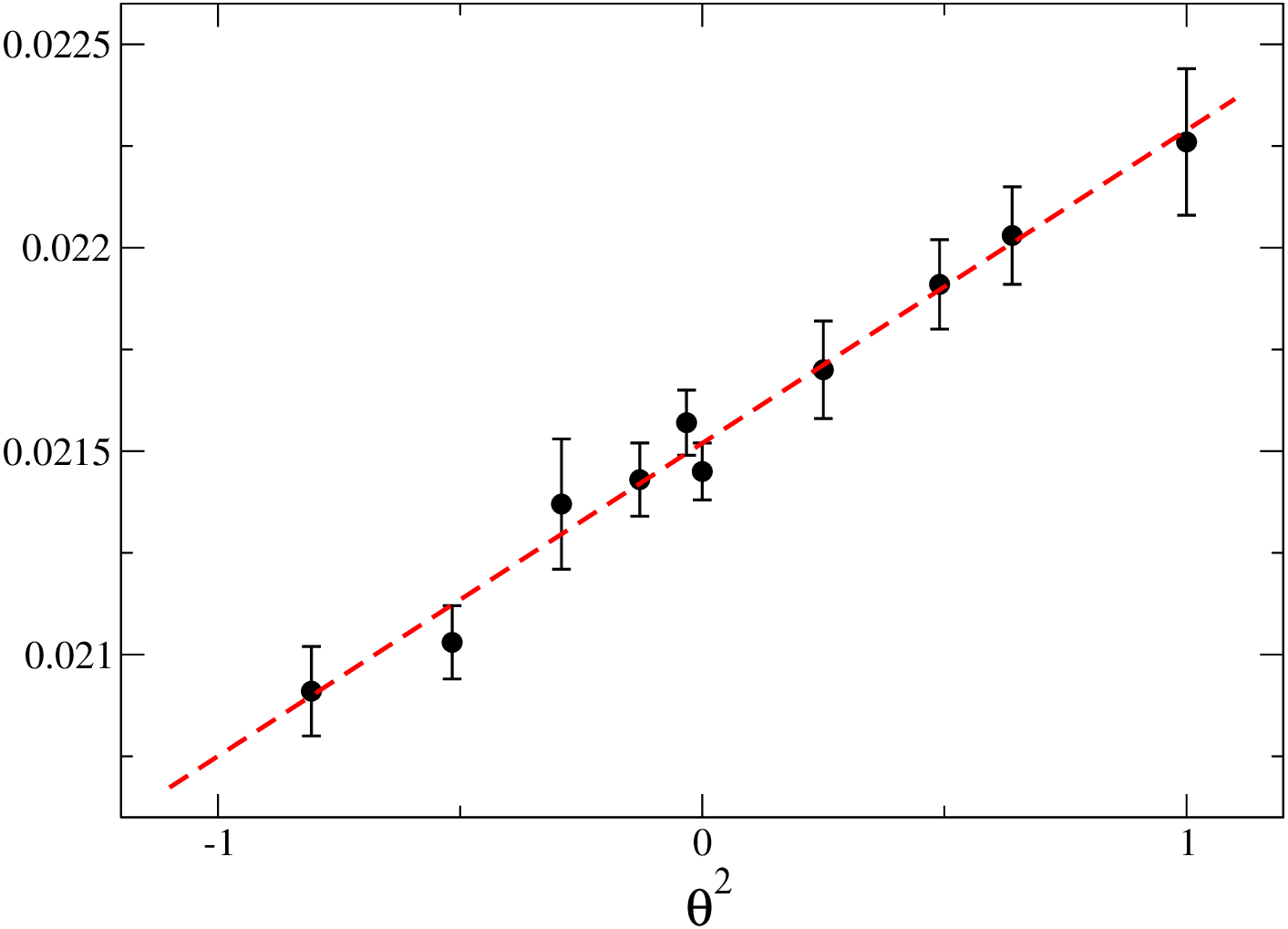}
\caption{Dependence of the Polyakov loop modulus on $\theta^2$ for 
$T \simeq 1.055\, T_c$ on the $40^3 \times 10$ lattice. The dashed line is a best fit according to a linear dependence on $\theta^2$.}
\label{fig:poltheta2}
\end{figure}

Let us now discuss the behavior of physical 
quantities computed at nonzero $\theta$ 
via reweighting, and compare it with results obtained at imaginary $\theta$.
We are interested, in particular, in the Polyakov loop modulus,
\beq
\langle |L| \rangle_\theta = \frac{\langle e^{i \theta Q} |L| \rangle_{}  }{\langle e^{i \theta Q} \rangle_{} } = \frac{\langle \cos(\theta Q) |L| \rangle_{} }{\langle \cos(\theta Q) \rangle_{} }
\label{formula:rewpoly}
\eeq
and in its susceptibility, 
$\chi_L (\theta) = 
V_s (\langle |L|^2 \rangle_\theta - \langle |L| \rangle^2_\theta)$.
The ratio of expectation values 
in Eq.~(\ref{formula:rewpoly}) 
is computed via a jackknife algorithm. We have replaced 
$e^{i \theta Q}$ with 
$\cos (\theta Q)$ also in the numerator, since $L$, as well as 
the path integral measure at $\theta = 0$, is invariant 
under parity transformations, under which instead $Q \to -Q$.

In Fig.~\ref{fig:poltheta2} we show the dependence 
of the Polyakov loop on $\theta^2$ for a selected value of the
bare coupling, $\beta=6.245$, corresponding to $T \sim 1.055\, T_c$.
Results at $\theta^2 \leq 0$ derive from direct simulations,
while those at $\theta^2 > 0$ have been obtained via reweighting from
$\theta = 0$ data. 
All data can be nicely fitted by a linear
dependence in $\theta^2$, as shown in the figure, demonstrating
that analyticity around $\theta = 0$ holds within errors.
The range of explored $\theta^2$ values is limited
on the right by the feasibility of reweighting, while on the left
one must avoid the crossing of the deconfining transition, which moves
to higher values of $T$ as $\theta^2$ decreases 
(see Fig.~\ref{fig:tctheta2}). 

The increasing behavior of
$\langle |L| \rangle$ can be understood considering that
one moves deeper and deeper into the deconfined phase as $\theta^2$ increases;
the quadratic behavior in $\theta^2$ is consistent with analyticity
around $\theta^2 = 0$ and with the fact that 
$\langle |L| \rangle$ is a $P$-even quantity.
We notice that both features are consistent with the results of 
Ref.~\cite{anber}.

\begin{figure}[t!]
\vspace{0.3cm}
\includegraphics*[width=0.47\textwidth]{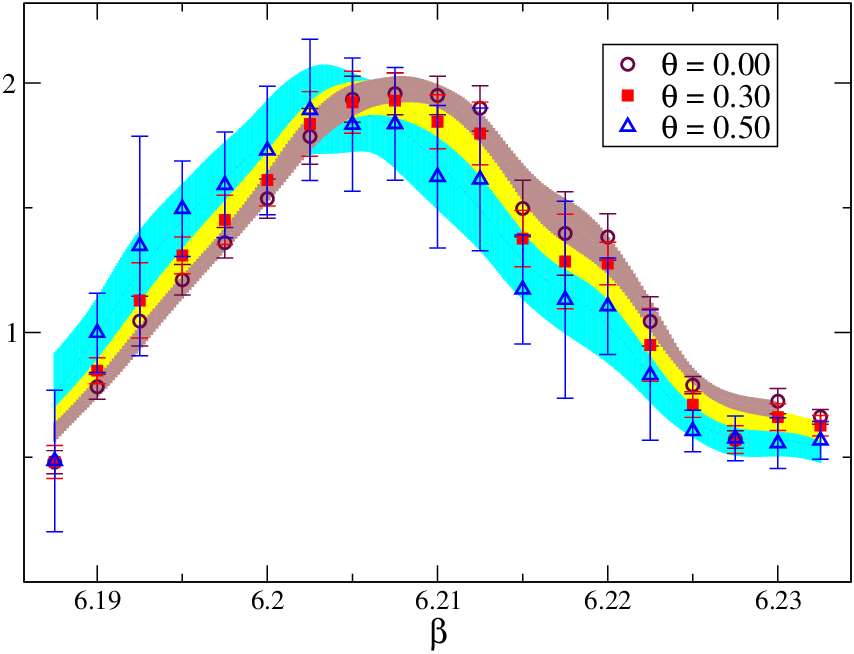}
\caption{Polyakov loop susceptibility as a function of $\beta$
and after reweighting at a few values of real $\theta$. 
The shaded bands correspond
to data reweighted also in $\beta$.}
\label{fig:suscrew}
\end{figure}

\begin{figure}[t!]
\vspace{0.46cm}
\includegraphics*[width=0.47\textwidth]{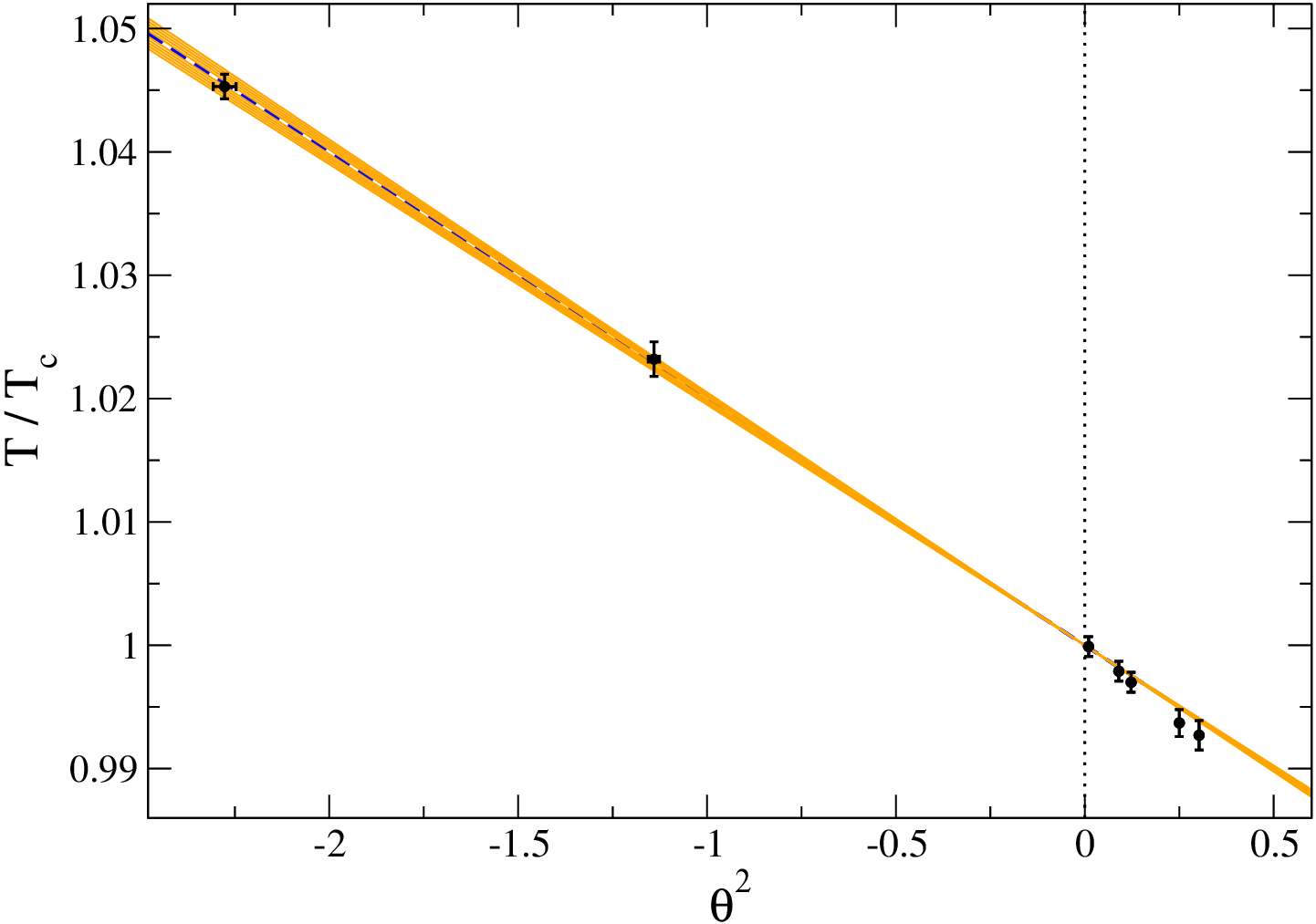}
\caption{Critical temperature as a function of $\theta^2$: we report the result
of the linear fit in $\theta^2$ obtained from simulations at $\theta^2 < 0$.}
\label{fig:tctheta2}
\end{figure}

Finally, in Fig.~\ref{fig:suscrew}, we show results for the
susceptibility as a function of $\beta$, obtained
after reweighting at $\theta = 0.3$ and $0.5$,
together with the original data at $\theta=0$.
It is clear that the peak moves to lower values of $\beta$, i.e. to
lower temperatures, as $\theta$ increases, in agreement with 
results from analytic continuation. From the susceptibility peaks we can 
extract the critical temperatures (see Table~\ref{tab:resrew}), 
and compare them with results at imaginary $\theta$. It does not make sense
to fit reweighted data directly, since they are obtained from the same 
data sample  and are therefore correlated; instead, in 
Fig.~\ref{fig:tctheta2}, we compare reweighted data with 
the extrapolation linear in $\theta^2$ obtained by fitting
results at imaginary $\theta$, showing that there is indeed agreement,
within statistical errors~\footnote{
If one wants to extract the curvature from reweighting, taking into account 
that results at different real values of $\theta$
are strongly correlated, 
then a reasonable estimate is obtained by considering 
only the point at the largest feasible value of $\theta$, 
i.e. $\theta = 0.55$. 
That yields, assuming that such a value is in the linear region, i.e. that
$R_\theta \simeq ( 1-T_c(\theta)/T_c(0))/\theta^2$, 
the value $R_\theta = 0.024(4)$,
in agreement with the estimate from analytic continuation
for $N_t = 10$, $R_\theta = 0.0200(5)$.}. That gives
further support to the validity of analytic continuation, at
least for small values of $\theta$.

\begin{table}
\begin{center}
\begin{tabular}{|c|c|c|c|}
\hline
lattice  & $\theta$ &  $\beta_c$ & $T_c(\theta)/T_c$ \\ \hline
\hline
$40^3 \times 10\ $  &   0.10\ & 6.2081(4)\   & 0.9999(8)\  \\
\hline
$40^3 \times 10\ $  &   0.30\ & 6.2068(4)\   &   0.9979(8)\     \\
\hline
$40^3 \times 10\ $  &   0.35\ & 6.2062(5)\   &   0.9970(8)\     \\
\hline
$40^3 \times 10\ $  &   0.50\ & 6.2040(6)\   &   0.9937(11)\     \\
\hline
$40^3 \times 10\ $  &   0.55\ & 6.2033(7)\   &  0.9927(12)\     \\
\hline
\end{tabular}
\end{center}
\caption{Results obtained for $\beta_c$ and $T_c$ at real $\theta$ by the reweighting technique on the $40^3 \times 10$ lattice. The ratios of critical temperatures have been calculated using the $\theta = 0$ critical $\beta$ reported 
in Table~\ref{tab:res}.}
\label{tab:resrew}
\end{table}

\subsection{Deconfinement and the Polyakov loop at 
fixed topological background.}
\label{fixed}

The general expression for a reweighted observable, 
\refeq{eq:rewgen}, can be rewritten in the following form:
\beq
\langle O \rangle_\theta = \frac{1}{\langle \cos(\theta Q) \rangle_{}}
\sum_{Q = -\infty}^{\infty} e^{i \theta Q}\, 
{\cal P} (Q)\,  \langle O \rangle_Q 
\label{eq:sector}
\eeq
where $\langle \cdot \rangle_Q$ stands for the average in a given
topological sector and ${\cal P} (Q)$ 
is the topological charge distribution
at $\theta = 0$. It shows that a non-trivial 
dependence on $\theta$ is possible only if the 
observable has a non-trivial dependence on $Q$.
This is quite natural, since $\theta$ and $Q$ are conjugate
quantities, like the particle density and the chemical potential.

The fact that, as we have shown, the location of deconfinement 
moves as $\theta$ is changed,
leads us to suspect
that the dependence of physical observables 
on $Q$ may be significant around $T_c$. 
Investigating such dependence is quite important
for various reasons, for instance to understand 
the possible systematic effects involved
in numerical simulations carried out in a fixed topological sector,
like it happens when investigating QCD with overlap fermions.
Studies regarding such effects have been reported, both at zero and finite 
$T$~\cite{brower,aoki07,cossu};
in particular, a recent study shows that
systematic effects in the determination of the topological 
susceptibility at finite $T$ are well under control~\cite{cossu}.
 In the present subsection 
we will discuss about the dependence on $Q$ of quantities directly
related to deconfinement, in particular the Polyakov loop and its
susceptibility, showing that in this case systematic effects, even if 
disappearing in the thermodynamical limit, can be more significant.

Such study is best
performed on the finest lattice at our disposal, 
 i.e. the $40^3 \times 10$, where the determination
of the topological background is most reliable. For that
reason we have divided the set of configurations sampled at 
each $\beta$ according to the value of $Q$ obtained via 
cooling, as discussed in the previous subsection. 
The expectation value $\langle \cdot \rangle_Q$ is obviously independent
of $\theta$ since, in a fixed topological background, 
$\theta$ only adds an irrelevant overall phase factor, hence in principle
one may think of combining equal $Q$ 
configurations sampled at different imaginary values of $\theta$. However,
one must consider that the lattice charge operator entering 
Eq.~(\ref{partfunlat}) contains irrelevant 
discretization terms, which are not constant over a given topological sector
and may lead to a residual dependence on $\theta_L$. For that reason,
in the following we will consider only configurations sampled at $\theta = 0$.

\begin{figure}[t!]
\vspace{0.46cm}
\includegraphics*[width=0.47\textwidth]{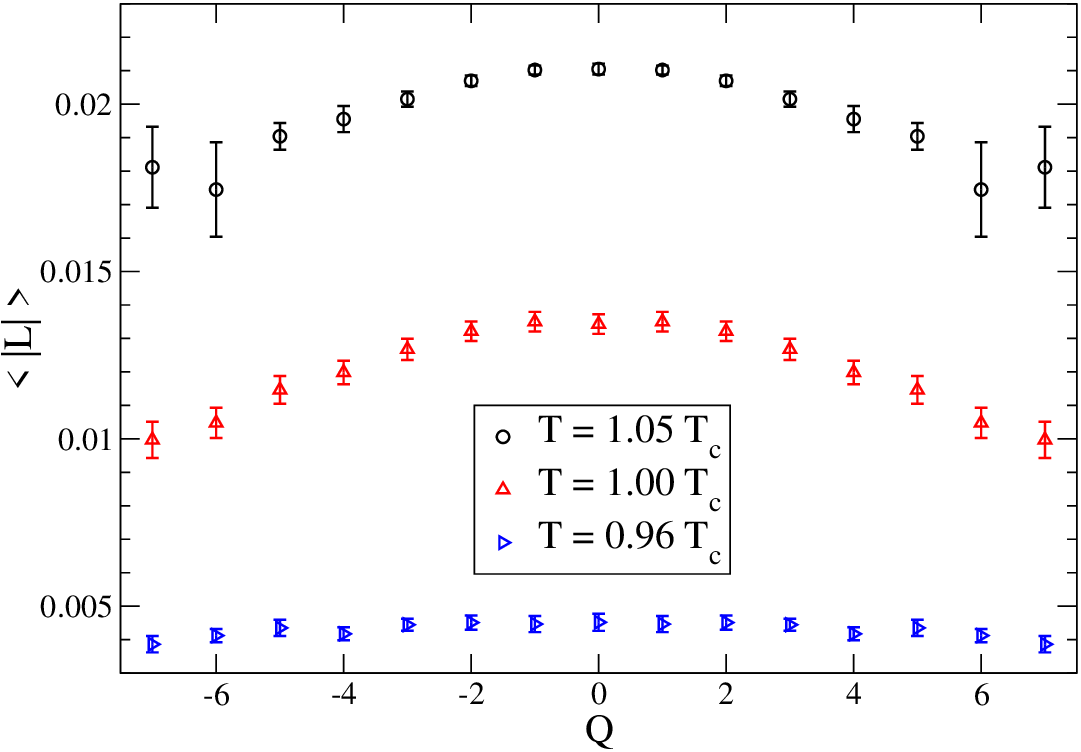}
\caption{Dependence of the Polyakov loop modulus on the topological 
sector $Q$, determined on the $40^3 \times 10$ lattice and 
for a few values of $T$ around the transition.}
\label{fig:poly_fixtopo}
\end{figure}

Let us start by showing, in Fig.~\ref{fig:poly_fixtopo},
 the behavior of the Polyakov loop 
as a function of $Q$ for a few temperatures around $T_c$
\beq
\langle |L| \rangle_{|Q|}=  
\frac{\sum_{i=1}^{N} |L|_i \,\,\delta_{|Q|,|Q_i|}}{\sum_{i=1}^{N} \delta_{|Q|,|Q_i|}} \, ,
\label{formula:polyfix}
\eeq
where $i$ runs over the $N$ measures and 
we have combined measures from opposite topological sectors,
exploiting the symmetry of the Polyakov loop under parity transformations,
in order to reduce statistical errors.
The exact symmetry visible in 
Fig.~\ref{fig:poly_fixtopo} is therefore artificial,
however we have verified that the symmetry holds, within errors, even 
before such combination.
We observe that, while 
below the transition the dependence on $|Q|$ is quite mild,
it gets stronger at the transition and becomes only slightly milder
above $T_c$. A similar behavior is observed for the average plaquette,
even if in this case the relative variation from one sector to the other
is always modest and never larger than $10^{-4}$.

\begin{figure}[t!]
\vspace{0.46cm}
\includegraphics*[width=0.47\textwidth]{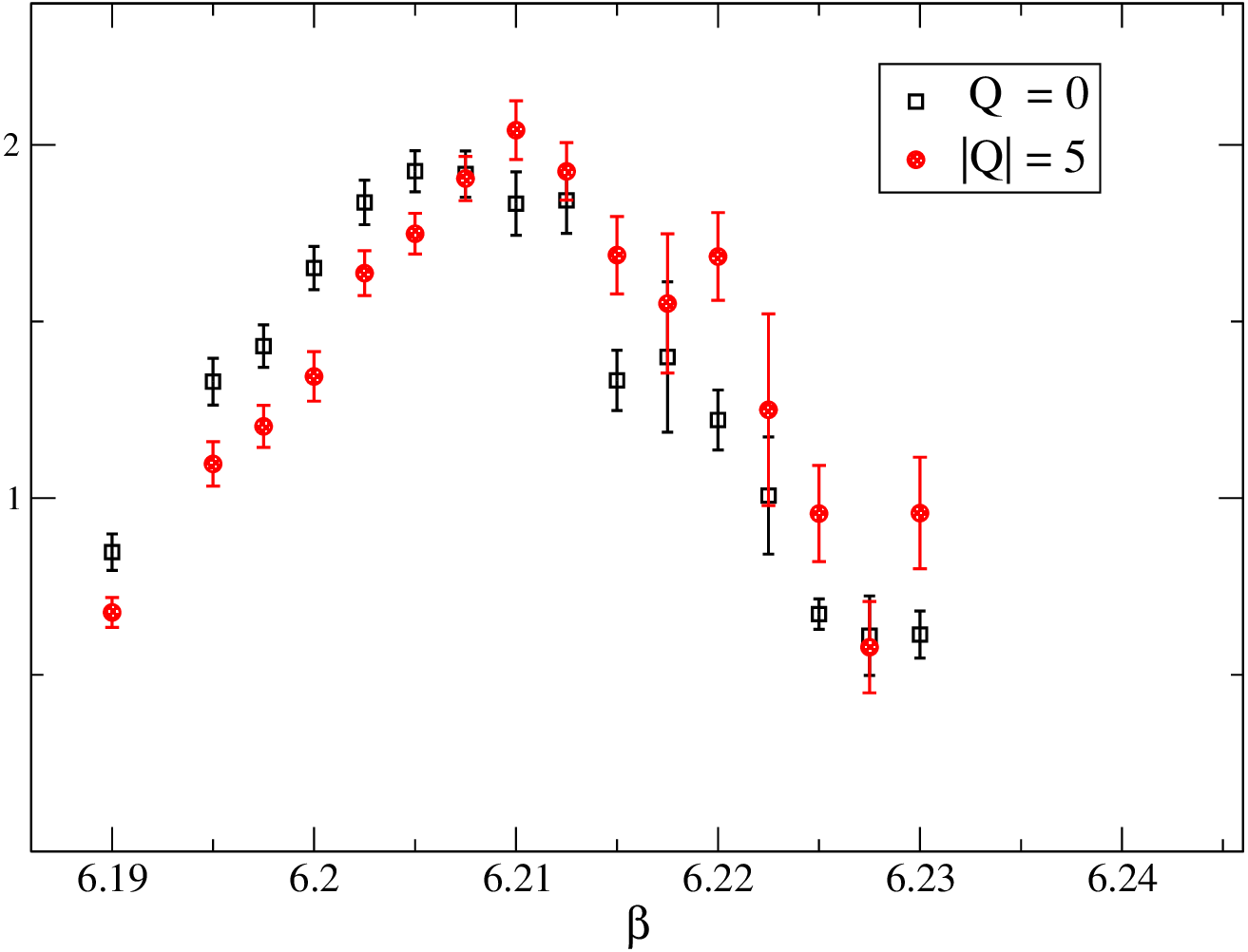}
\caption{Susceptibility of the Polyakov loop as a function of $\beta$
on the $40^3 \times 10$ lattice, determined after fixing 
the topological sector.}
\label{fig:susc_fixtopo}
\end{figure}

The dependence on $Q$ is quite visible also in the 
susceptibility of the Polyakov loop,
which is shown
in Fig.~\ref{fig:susc_fixtopo} as a function of $\beta$ for 
$Q=0$ and $|Q|=5$. The shift of the susceptibility peaks tells
us that even the transition temperature can be influenced
by the overall topological background. In particular,
in Table~\ref{tab:resfixtopo}, we report the values of  
$T_c(Q)$, obtained by fitting such peaks with Lorentzian functions.
The critical temperature tends to increase as $|Q|$ increases; this is 
qualitatively consistent with what found when adding an imaginary $\theta$ 
term, which has the effect of shifting the average value
of the topological charge distribution towards non-zero values.

\begin{table}
\begin{center}
\begin{tabular}{|c|c|c|c|}
\hline
lattice  & $|Q|$ &  $\beta_c$ & $T_c(Q)/T_c$ \\ \hline
\hline
$40^3 \times 10\ $  &   0\ & 6.2065(5)\   & 0.9975(10)\  \\
\hline
$40^3 \times 10\ $  &   1\ & 6.2068(5)\   &   0.9978(10)\     \\
\hline
$40^3 \times 10\ $  &   2\ & 6.2069(5)\   &   0.9981(10)\     \\
\hline
$40^3 \times 10\ $  &   3\ & 6.2080(5)\   &   1.0000(10)\     \\
\hline
$40^3 \times 10\ $  &   4\ & 6.2092(5)\   &  1.0015(10)\     \\
\hline
$40^3 \times 10\ $  &   5\ & 6.2108(7)\   &  1.0039(12)\     \\
\hline
$40^3 \times 10\ $  &   6\ & 6.2118(7)\   &  1.0053(12)\     \\
\hline
\end{tabular}
\end{center}
\caption{Results obtained for $\beta_c$ and $T_c$ at fixed topology calculated with $\beta_c=6.2082(4)$.}
\label{tab:resfixtopo}
\end{table}

\begin{figure}[t!]
\vspace{0.46cm}
\includegraphics*[width=0.47\textwidth]{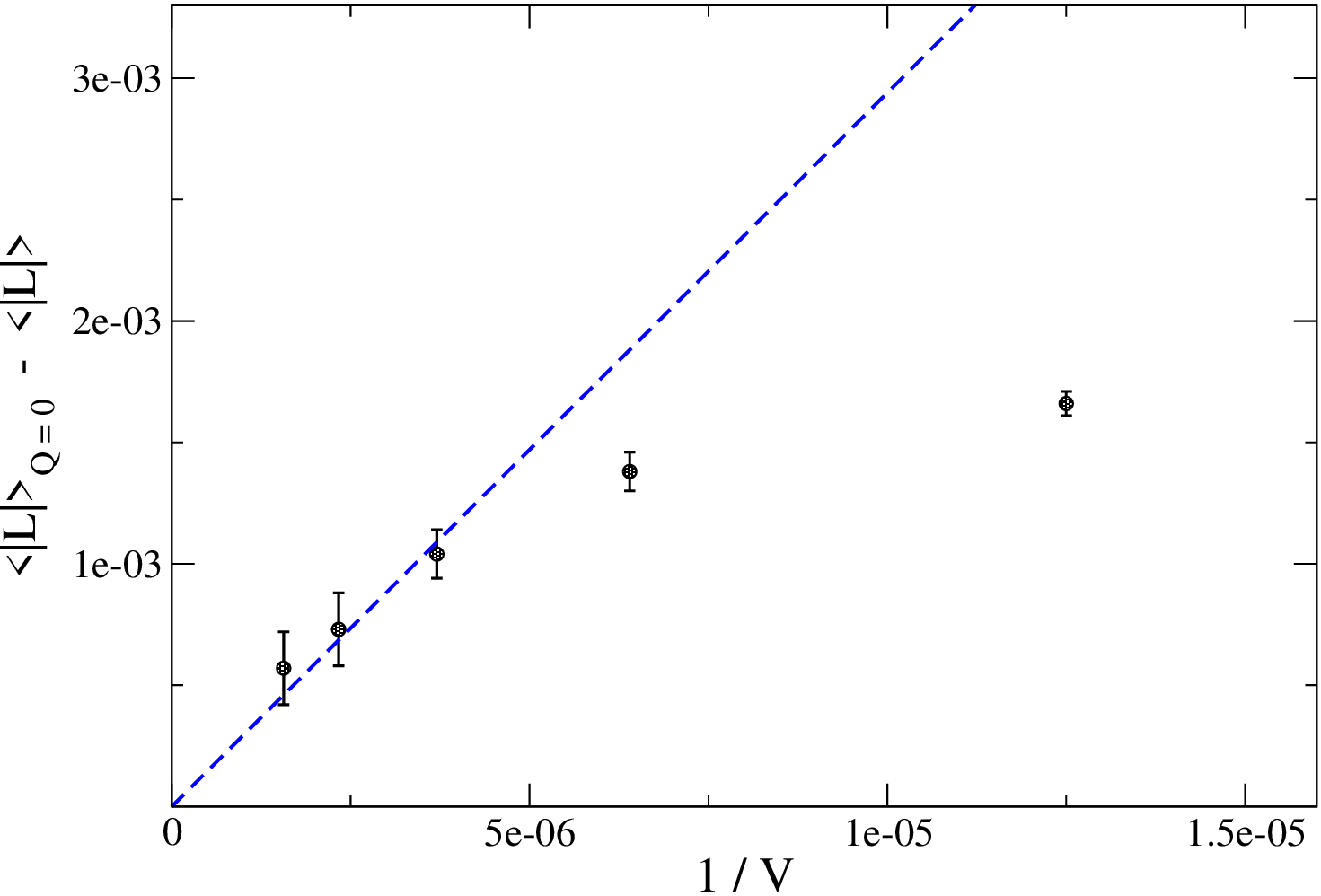}
\caption{Variation of the Polyakov loop modulus in the 
$Q = 0$ sector, with respect to the average over all sectors, 
plotted as a function of ${1}/{V}$,
for $T \simeq 1.018\, T_c$. The dashed line is the result of a linear
fit in $1/V$.}
\label{fig:loop0vol}
\end{figure}

One expects that systematic effects present in a fixed sector $Q$ disappear
as the thermodynamical limit is approached. 
In order to verify that, we have performed, for 
a given value of $T \simeq 1.018\, T_c$ ($\beta = 6.22$), 
simulations on lattices with different spatial volumes
($L_s^3 \times L_t$ with $L_t = 10$ and $L_s = 16,\, 18,\, 20,\, 25,\, 30,\,
35,\, 40$), then combining measures obtained within different
topological sectors $Q$ as described above. 
In Fig.~\ref{fig:loop0vol} we show how the difference of the Polyakov 
loop modulus in the $Q = 0$ sector, taken with respect to its average over all
sectors, changes as a function of the volume $V = L_t\, L_s^3$. 
The difference clearly approaches zero linearly in $1/V$, as one 
indeed expects on general grounds.

We will now try to better describe the observed dependence
of the Polyakov loop on $Q$ by a very simplified model, 
which is based on the instanton gas approximation and
follows the analysis reported in Ref.~\cite{brower}. Let us consider
a generic, extensive quantity, like the average Polyakov loop times
the volume $V$: we assume that it receives a given, fixed contribution
by each topological object, instanton or anti-instanton, and that
the topological objects are distributed according to the 
instanton gas approximation, i.e. that the probability 
of having $n$ instantons and $\bar n$ anti-instantons is given by
\beq
{\cal P} (n,\bar n) = e^{-2 \lambda}\,  
\frac{\lambda^n \lambda^{\bar n}}{n !\, \bar n !}
\label{poisson}
\eeq
where $2 \lambda = \langle Q^2 \rangle = V\, \chi_l$
and $\chi_l = a^4\, \chi$. 
The relevant quantity, to describe the behavior as a function of
$Q = n - \bar n$, is the average of the total number of topological objects
which are found  at fixed $Q$, $\langle n + \bar n \rangle_{Q}$, which 
can be extracted as a constrained average starting from the double
Poissonian distribution in Eq.~(\ref{poisson}). The result obtained
at the lowest order in $Q^2 / (2 \lambda) = Q^2 / \langle Q^2 \rangle$, which is the relevant expansion
parameter when approaching the thermodynamical limit, is~\cite{brower}
\beq
\langle n + \bar n \rangle_Q \simeq 
2 \lambda 
-\frac{1}{2} \left( 1 - \frac{Q^2}{2 \lambda} \right)\, .
\label{expansion}
\eeq
The prediction for $\langle |L| \rangle_Q$, 
which follows from our simplified model, is then
\beq
\langle |L| \rangle_Q &=& {\rm const} 
- \frac{\gamma}{V} \langle n + \bar n \rangle_Q \nonumber \\
&\simeq& 
\langle |L| \rangle 
+ \frac{\gamma}{2 V} \left( 1 - \frac{Q^2}{V\, \chi_l} \right) \, .
\label{loopq}
\eeq
where we have defined as $-\gamma$ the contribution to $V |L|$ coming 
from each (anti)instanton and we have exploited the fact that
the expression in parentheses vanishes when taking the average 
over all sectors. 

\begin{figure}[t!]
\vspace{0.46cm}
\includegraphics*[width=0.47\textwidth]{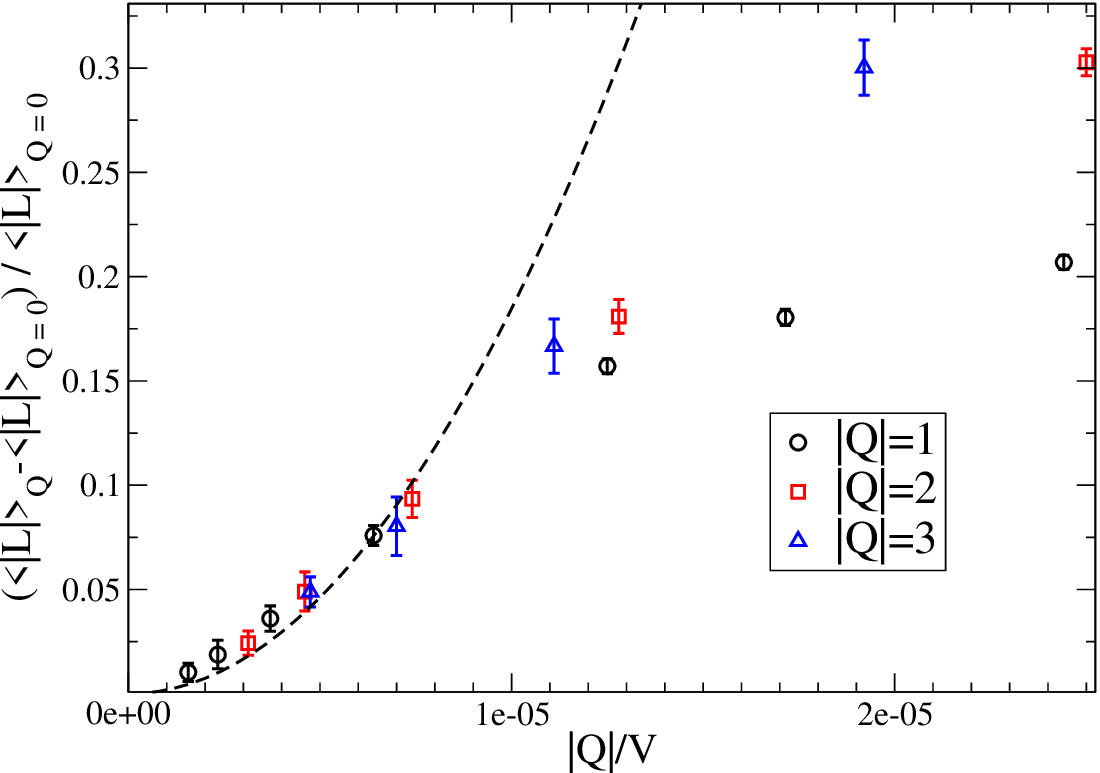}
\caption{Relative variation of the Polyakov loop modulus with 
the topological background, plotted 
as a function of the topological charge density $\frac{Q}{V}$,
for $T \simeq 1.018\, T_c$.
Results have been obtained on different spatial volumes, the dashed
line is the prediction from the simple model described in the text.}
\label{fig:finitevol}
\end{figure}

Eq.~(\ref{loopq}), which is expected to be valid
as the thermodynamical limit is approached,  predicts 
$\langle |L| \rangle_Q - \langle |L| \rangle$ to vanish linearly
in $1 / V$. This is confirmed by the behavior shown in Fig.~\ref{fig:loop0vol},
and a linear fit to data on the larger volumes, which is shown in
the same figure, gives back 
$\gamma \sim 6 \times 10^2$. It is interesting that, once fixed $\gamma$ and 
knowing from the average over the whole ensemble 
that $\chi_l = \langle Q^2 \rangle / V \sim 0.947 \times 10^{-5}$, the behavior
of the Polyakov loop as a function of $Q$ in the large volume limit
is completely fixed by the model, in particular 
$\langle |L|\rangle_{Q=0} - \langle |L|\rangle_{|Q|} 
\simeq \gamma Q^2/ (2 \chi V^2)$. In order to check that, 
in Fig.~\ref{fig:finitevol} we plot the quantity
\beq
\Sigma(|Q|)= \frac{\langle |L|\rangle_{Q=0} - \langle |L|\rangle_{|Q|} }{\langle |L|\rangle_{Q=0} }\, ,
\label{formula:polyratio}
\eeq
which gives the relative deviation of the Polyakov loop from the value
it takes in the trivial topological sector (the error on 
$\Sigma(|Q|)$ has been obtained by a jackknife algorithm).
In particular, we plot $\Sigma(|Q|)$ as a function of $|Q|/V$ for 
$|Q|=1,\,2,\,3$ and for all the explored volumes, together
with the model prediction, which has no more free parameters left.
The fair agreement observed for small values of $|Q|/V$ is therefore
highly non-trivial, given the crudeness of the model: part of the success 
can be ascribed to the rapid approach to the instanton gas approximation which
takes place right above $T_c$, as demonstrated by the results
of Ref.~\cite{BDPV}. As $|Q|/V$ increases, however, the topological background
is not dilute enough and the model prediction fails.

It would be nice to study the interplay between topological activity and the holonomy
in more detail, in particular approaching the deconfining transition from above, 
and compare with model studies about the same issue (see, e.g., Ref.~\cite{shsu13}),
however that goes beyond the purpose of our present investigation.

Finally, it is important to stress that,
despite the fact that the approach to the thermodynamical limit of 
$\langle |L| \rangle_Q$ seems to be well understood and that systematic
effects vanish as $1/V$, from Fig.~\ref{fig:finitevol}
we learn that they are still appreciable,
and of the order of $10\%$, even on the largest explored volume,
whose aspect ratio  $L_s/L_t = 4$ is common to many 
finite temperature computations found in the literature.

\section{Phase diagram in the $T-\theta$ plane: 
general features and analogies with the diagram at imaginary $\mu_B$}
\label{phdiagram}

After studying $T_c(\theta)$, it is tempting to draw a sketch of the 
whole phase diagram in the $T$-$\theta$ plane. On the imaginary 
side, $\theta = i\, \theta_I$, no particular structure is expected apriori, since
$CP$ symmetry is explicitly broken whenever ${\rm Im} (\theta) \neq 0$.
Indeed, we have not observed any transition, apart from the 
deconfining one, in the range of explored values of $\theta_I$, even if 
we cannot exclude the presence of new phase structures at 
larger values of $\theta_I$.

The situation is quite different for real $\theta$, which plays the role
of an angular variable. Periodicity in $\theta$ must reflect in some
way in the structure of the phase diagram in the $T$-$\theta$ plane, which
is then expected to be non-trivial. It is interesting to notice
that this is very similar
to what happens in presence of an imaginary baryon chemical potential
$\mu_B$, and indeed many analogies can be found between 
the $T$-$\theta$ phase diagram and the phase structure at imaginary
$\mu_B$~\cite{agl03,kouno1,kouno2,kouno3,dene}.
It is convenient, for the following discussion, to introduce
the parameter $\theta_B \equiv {\rm Im}(\mu_B)/T$, since 
in terms of it analogies appear more clearly.

The purpose of the present Section is to discuss such analogies, also
in a large $N$ perspective, with a particular emphasis on  duality,
in the sense of an inversion between the high and low temperature regions,
between the $T$-$\theta$ and $T$-$\theta_B$ phase diagrams,
which can be
suggestive of the possible dual role played by the respective 
relevant degrees of freedom.

We will start by
giving a rapid overview about the $T$-$\theta_B$ phase diagram, in order 
to highlight aspects which may have a direct correspondence 
with the case of the $T$-$\theta$ plane, which is discussed afterwards.

\subsection{Phase diagram in the $T$-$\theta_B$ plane}

Let us consider QCD with $N$ colors and
its partition function at non-zero baryon chemical potential,
\beq
Z(T,\mu_B) = {\rm Tr} \exp \left( -\frac{H - \mu_B B}{T} \right) \, ,
\eeq
where $H$ is the QCD Hamiltonian and $B$ is the baryon number operator.
For purely imaginary values of $\mu_B$, which are often considered to avoid 
the sign problem, the partition function becomes
\beq
Z(T,\theta_B) = {\rm Tr} \left( e^{-\frac{H}{T}}\ e^{i \theta_B B} \right)
\eeq
where $\theta_B = {\rm Im}(\mu_B)/T$. 

It is clear that 
$\theta_B$ plays the role of an angular variable, however the actual 
dependence of the free energy on $\theta_B$ depends 
on the phase  of the theory. In the confined phase, 
$\theta_B$ couples only to physical degrees of freedom 
which have integer baryon charge
$B$, hence the free energy is a function of $\theta_B$ with period $2 \pi$.
In the deconfined phase, instead, new physical degrees of freedom appear,
quarks, 
carrying a fractional baryon charge, in particular 
in units of $1/N$: as a consequence
the free energy is expected to be a function of $\theta_B/N$.

One may expect then that the periodicity in $\theta_B$ be $2 \pi N$,
but instead  it is easy to prove that, independently of the relevant degrees
of freedom, the partition function must be periodic in $\theta_B$ with period 
$2 \pi$. Indeed, in the path integral representation of the partition
function
\beq
Z = \int \mathcal{D} A e^{-S_{G}[A]} \det M[A]\, ,
\eeq
where $\det M[A]$ is the quark determinant,
the imaginary chemical potential enters as a twist, by a phase factor
$\exp( i \theta_B/N)$, in the boundary conditions for quark fields. 
However, for $\theta_B = 2 \pi k$, with $k$ integer, such twist can 
be cancelled exactly by a center transformation on the gauge fields
i.e. by a gauge transformation periodic in time up to a global
element of $Z_N$, the center of the gauge group, $\exp(-2 \pi k/N)$, under
which the pure gauge action is invariant. As a 
consequence, the partition function and the free energy
must be always periodic in $\theta_B$,
with period $2 \pi$.

How is it possible to reconcile such periodicity with
the expected dependence on $\theta_B/N$ in the deconfined phase? 
This is done by a non-analytic, multi-branched 
behavior of the free energy, as a function
of $\theta_B$, in the high temperature deconfined phase, 
with phase transitions happening at $\theta_B^{(RW)} = (2 n + 1) \pi$ ($n$ 
being a relative integer)  and known as 
Roberge-Weiss (RW) transitions~\cite{rw}.
 When crossing such values, gauge fields
jump discontinously from one center sector to the other, characterized
by a different global alignment of the Polyakov loop.
One has, therefore, $N$ different branches, which are not equivalent
from the point of view of the order parameter, $\langle L \rangle$, 
but whose free energies are identical, modulo a shift 
$\theta_B \to \theta_B + 2 \pi$, by virtue of the invariance of 
the pure gauge action under center transformations.

Let us try to better clarify the role played by center symmetry.
$Z_N$ is broken explicitly by the presence 
of the quark determinant; however, a residual $Z_2$ symmetry exists for
particular values of $\theta_B$, $\theta_B = k \pi$, with
$k$ relative integer. Such residual symmetry
can be identified, modulo a phase rotation, with charge conjugation $C$.
It stays always unbroken for even values of $k$; on the contrary,
it breaks spontaneously, in the high $T$ phase, 
for odd values of $k$, for which the effective potential of the 
Polyakov loop has two equivalent, degenerate minima, 
corresponding to adiacent center sectors~\cite{rw}.

\begin{figure}[t!]
\includegraphics*[width=0.47\textwidth,height=0.30\textwidth]{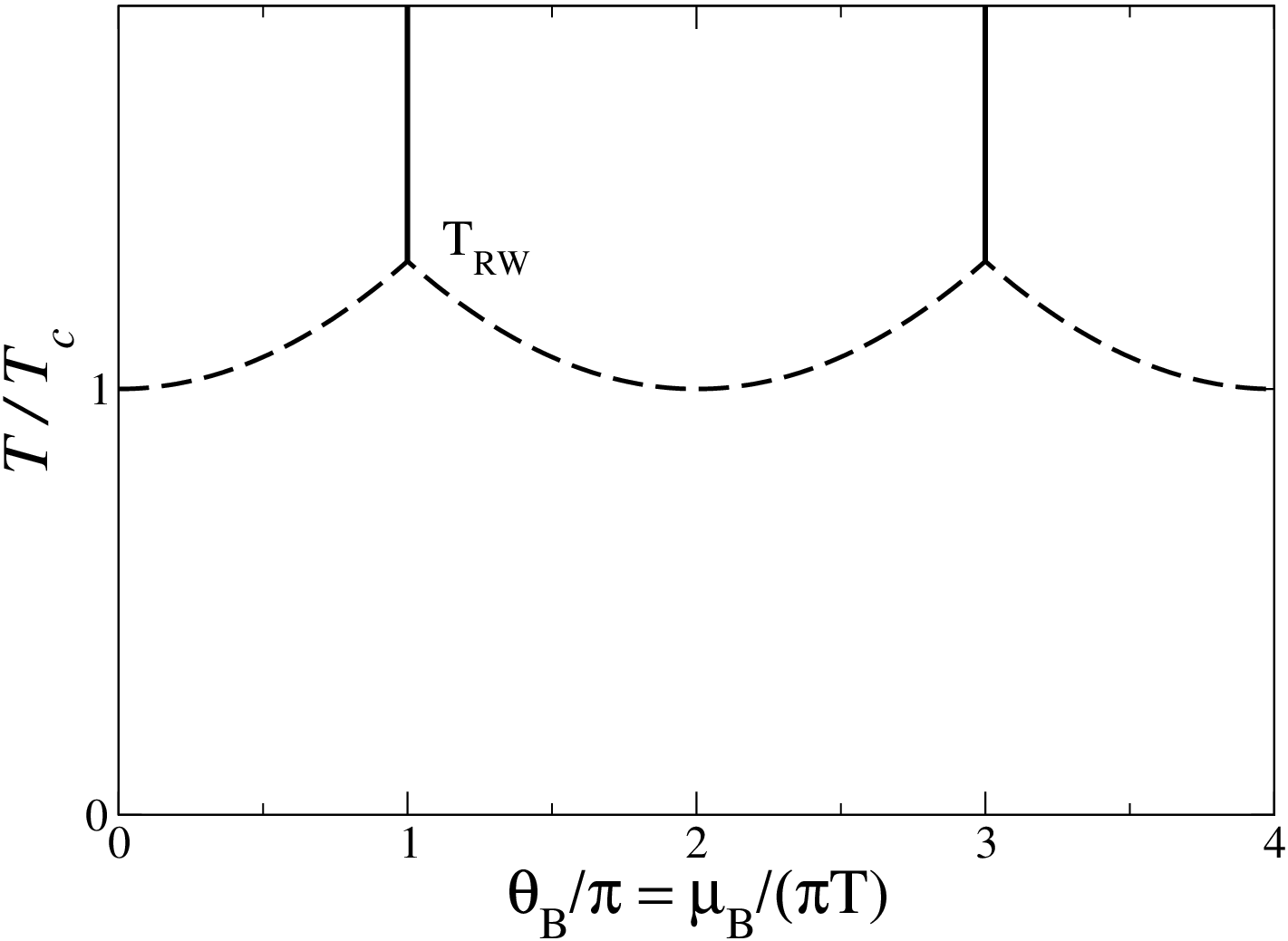}
\caption{Phase diagram of QCD in presence of an imaginary baryon chemical 
potential $\mu_B$, as it emerges from symmetry considerations and numerical
simulations. The vertical lines are the Roberge-Weiss transition lines present in the high-$T$ phase of the theory, the dashed lines represent the 
deconfining transition and $T_{\rm RW}$ indicates the temperature at which the Roberge-Weiss lines terminate (see text).}
\label{fig:thetabdiag}
\end{figure}

The RW transitions and their connection with the deconfining,
chiral restoring (pseudo)-critical line $T_c(\theta_B)$ have been
widely studied both by numerical lattice simulations
and by effective model 
computations\cite{sqgp,rwep,rwep_nf3,rwep2,rwep3,kouno,braun,sakai,aarts,pawlowski,rafferty,pagura,kashiwa,morita,nagata,wu}. The resulting
diagram in the $T$-$\theta_B$ plane is sketched in Fig.~\ref{fig:thetabdiag}.

The RW transition lines are first order and correspond to a discontinous jump
in the order parameter, the Polyakov loop. The order of their endpoint,
instead, depends on the quark mass spectrum: evidence from lattice studies
collected up to now is that, both for the two-flavor and the 
three-flavor theory, the endpoint is second order 
for intermediate quark masses
and first order in the limit of large or small quark masses~\cite{rwep,rwep_nf3,rwep2}. 
In the former 
case the universality class is that of the 3D Ising model,
since the relevant symmetry is $Z_2$; in the latter,
the endpoint is actually a triple point, with two further first
order lines departing from it, which can be identified with part
of the (pseudo)critical lines $T_c(\theta_B)$ corresponding to chiral symmetry
restoration and deconfinement. The line $T_c(\theta_B)$ is therefore
a multibranched function itself, with cusps which can be conjectured 
to coincide with the RW endpoints, as depicted
in Fig.~\ref{fig:thetabdiag}: that is also consistent with available 
numerical evidence.

\subsection{Phase diagram in the $T$-$\theta$ plane}

\begin{figure}[t!]
\includegraphics*[width=0.47\textwidth,height=0.30\textwidth]{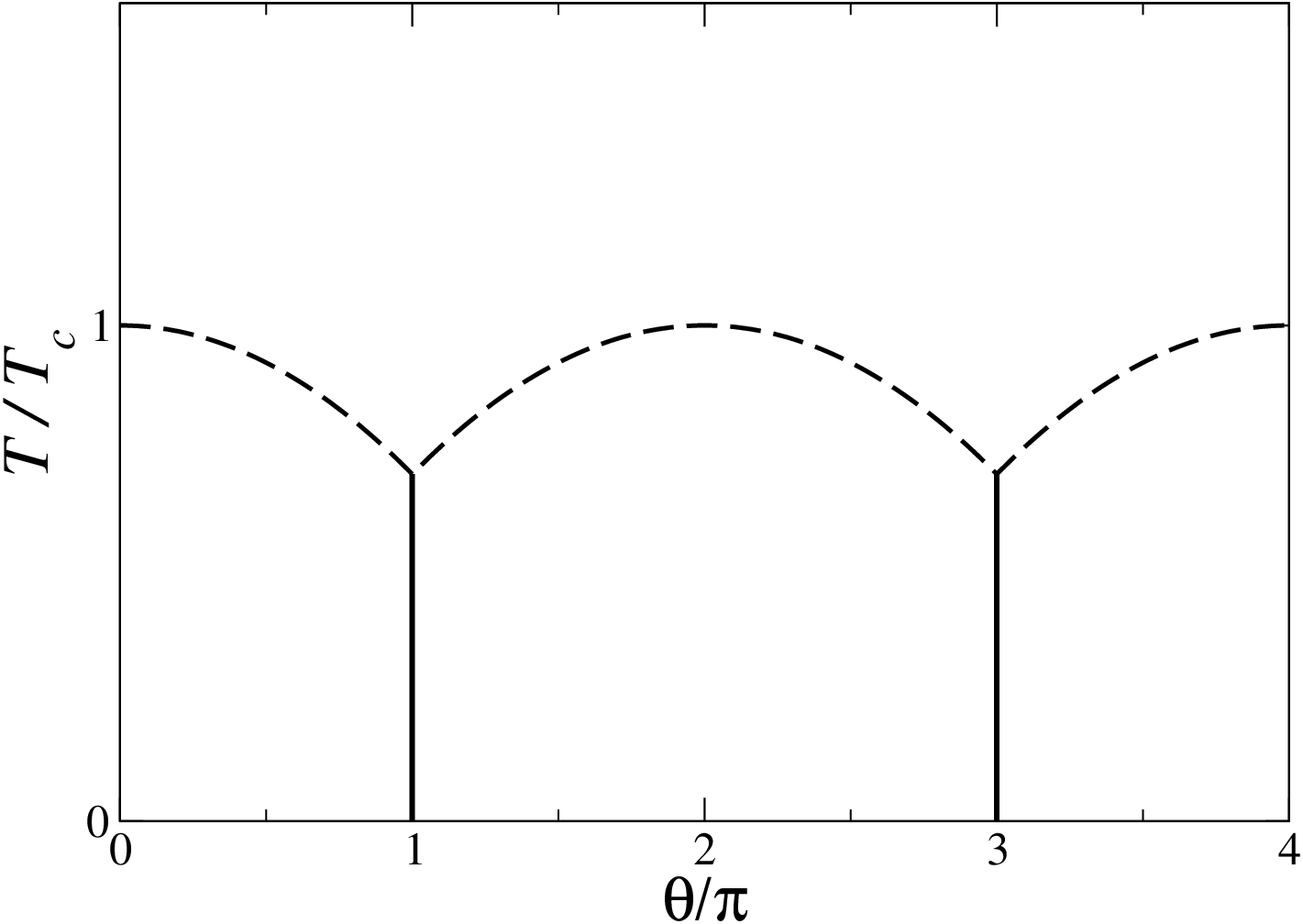}
\caption{Conjectured phase diagram of the pure gauge $SU(N)$ Yang-Mills 
theories in the $T$-$\theta$ plane. 
The vertical lines are the first order transition lines expected in the 
low-$T$ phase of the theory, the dashed lines correspond to the deconfining
transition.}
\label{fig:thetadiag}
\end{figure}

In presence of a real $\theta$ term, gauge configurations are weighted,
in the path integral representation
of the partition function, 
by a factor $\exp (i \theta Q)$. 
The topological charge $Q$ is globally integer
for finite action configurations, hence the partition function
and the free energy must be periodic in $\theta$, with period $2 \pi$.

However, if one searches for a $\theta$ dependence which 
stays non-zero at the leading order in $1/N$, 
as required by the solution
to the axial $U(1)$ problem, 
one needs that the free energy
be a function of $\theta/N$, instead of $\theta$,
otherwise $\theta$ dependence would be suppressed
exponentially in $N$~\cite{u1wit}.

Also in this case, the only
possible way to reconcile periodicity in $\theta$ and 
dependence on $\theta/N$ is to admit that the free energy density
$f(\theta)$ is a multibranched function of $\theta$~\cite{wit80,ohta,wit98}, 
scaling in the large
$N$ limit as follows~\cite{wit80,wit98}:
\beq
f(\theta) = N^2 {\rm min}_k \, h \left(\frac{\theta+2\pi k}{N} \right)
\label{largenscal}
\eeq
where $k$ runs over all relative integers. For each value of $\theta$
the system choices the branch which minimizes the free energy. 
The function $h$ can be chosen so as to have 
its minimum in zero~\cite{vafawit}, 
so that the branch relevant to $\theta \sim 0$ 
corresponds to $k = 0$. Moreover, the invariance under $CP$, present
at $\theta = 0$, imposes that $h$ is an even function of $\theta$.   

A shift $\theta \to \theta + 2 \pi$ corresponds to a passage from one 
branch to the other, which, according 
to the large $N$ scaling in Eq.~(\ref{largenscal}), must 
happen discontinously,
in points where the free energies of the two adiancent branches cross with
different (opposite) derivatives, i.e. through a first order transition. 
For symmetry 
reasons that happens for $\theta = \pm \pi$, or odd
multiples of such values. $CP$ symmetry, which is exact in correspondence
of such points, is broken spontaneously by the choice of 
one of the two equivalent branches, which are not invariant under
$CP$, but instead exchanged into each other. 
$CP$ is of course exact also for $\theta = 0$ and for 
integer multiples of 
$2 \pi$, but there no spontaneous breaking happens.

The scenario depicted above is true only for sufficiently 
low temperatures. Indeed, in the opposite limit of high $T$, the 
instanton gas approximation must set in, which predicts a smooth,
periodic behavior in $\theta$, but with an exponential suppression
in the limit of large $N$~\cite{CDG-78,GPY-81}~\footnote{
The same approximation does not work in the low $T$ region, because
of infrared divergences. 
}.

Actually, it has been 
conjectured~\cite{ztn1,wit982,saksug,bergman,ztn2,ztn2b,ztn3} 
and recently proven by lattice
simulations~\cite{BDPV} that the change in $\theta$ dependence happens
exactly in correspondence of the deconfinement transition. Therefore, 
while the confined
phase is characterized by a dependence on $\theta/N$ and a non-analytic 
periodicity in $\theta$, induced by the multi-branched structure of the 
free energy, the deconfined phase is characterized
by a smooth, periodic dependence on $\theta$, but suppressed like
$e^{-N}$ in the large $N$ limit, which very rapidly approaches the 
instanton gas prediction\footnote{Evidence from Ref.~\cite{BDPV}, extracted by looking at higher order cumulants of the topological charge distribution, is that the 
instanton gas approximation sets in around $T \sim 1.1\ T_c$
for the $SU(3)$ pure gauge theory.
}: in this case, of course, no
$CP$ breaking transition is expected at $\theta = \pi$.
Hence, the $CP$ breaking transition lines, present for 
$\theta^{(CP)} = (2n + 1) \pi$ and for low enough temperatures, must 
end at some temperature around $T_c$: 
as for the case of the RW lines, since the relevant symmetry
is again $Z_2$, their endpoint can be either
first order or second order in the 3D Ising universality class.
\\

The scenario described above is reproduced in Fig.~\ref{fig:thetadiag},
where we have drawn a sketch of the $T$-$\theta$ phase diagram. We have 
not made any discussion yet regarding the deconfinement line, $T_c(\theta)$, but let
us stop for while to comment on
the analogy with the periodic structure in the $T$-$\theta_B$ plane,
which now appears quite clearly. The analogy actually implies an exchange
between the high $T$ and the low $T$ regions of the two diagrams, 
i.e. $T \to 1/T$, which
is suggestive of the possible dual role played by the relevant degrees of 
freedom in the two cases.

We have already discussed how the analytic or non-analytic periodic
structure of  the $T$-$\theta_B$ can be understood in terms of 
a dependence of the free energy on $\theta_B$ (low $T$) or 
$\theta_B/N$ (high $T$), which in its turn stems from the relevant degrees
of freedom being hadrons with integer baryon charge $B$, or quarks carrying
fractional baryon charge in unit of $1/N$. If one wants to apply a similar,
intuitive picture to the $T$-$\theta$ plane, one has to conjecture
that the relevant topological degrees of freedom carry integer values
of $Q$ in the high $T$, deconfined phase 
(in agreement with the instanton gas picture), 
but instead fractional charges, in units of $1/N$, in the low $T$,
confined phase. 

Actually, such hypothesis is not new. Indeed, the possible 
existence of topological objects with fractional charge,
which sometimes are  called instanton quarks, and their possible 
role in the confined phase,
have been conjectured since long~\cite{wit78,fateev,berglu,belavin,unsal08,diakonov,gorsky,ztn2,ztn2b,ztn3,kraan,bruck03,bruck04,dunne}.
In the high $T$, deconfined phase, 
they are expected to be localized and confined into 
larger objects carrying integer topological charge, like instantons 
and calorons. Instead in the low $T$, confined phase, they are 
expected to be free and delocalized topological objects.\\

Apart from the $CP$ breaking lines, the conjectured diagram
sketched in Fig.~\ref{fig:thetadiag}
is completed by the deconfinement line, $T_c(\theta)$, that we have 
discussed for $SU(3)$ and small values of $\theta$ in the previous 
Section. It is reasonable to assume, also based on the large $N$
model of Ref.~\cite{dene}, that
$T_c(\theta)$ is fixed by the interplay between thermodynamics
and topological properties of the theory: since topology is exponentially
suppressed in the deconfined phase, the leading dependence of $T_c(\theta)$
must derive from the topological properties of the confined phase,
hence we expect~\cite{dene} 
that also $T_c(\theta)$ be a multibranched function
of $\theta/N$, dominated, at the leading order in $1/N$, by the quadratic
term
\beq
T_c(\theta)/T_c(0) \simeq 1 - R_\theta 
\min_k\left(\theta+2\pi k\right)^2
\eeq
where $k$ is a relative integer and $R_\theta$
is $O(1/N^2)$. 

Therefore, our expectation is that, at least for large 
enough $N$, the deconfinement temperature tends to a finite, non-zero
value at $\theta = \pi$ or odd multiples of it 
(actually, $\theta$ independence is expected as $N \to \infty$).
Periodicity in $\theta$ implies the presence of 
cusps for $T_c(\theta)$ at $\theta = (2 k + 1)\, \pi$. 
Notice that the phase structure may be more complicated 
in presence of dynamical fermions,
which have not been considered
in this context: one can still predict the presence of a zero temperature,
CP breaking transition at $\theta = \pi$~\cite{dashen,divecchia,smilga,tytgat,creutz},
however the interplay between chiral symmetry breaking and confinement can
lead to a richer diagram, with the possibility of considering 
also quark chemical potentials or external 
fields~\cite{metlit,fraga,kouno1,kouno2,kouno3,boomsma1,boomsma2}.

There is one feature of the sketch in Fig.~\ref{fig:thetadiag} which is pure 
speculation, stimulated by the analogy with the $T$-$\theta_B$ plane:
the curve $T_c(\theta)$ hits
the $CP$ breaking lines exactly at their endpoints, as it happens 
for $T_c(\theta_B)$ with the RW lines. If such speculation were correct,
then, since $T_c(\theta)$ is first order at least for large $N$, 
it would be reasonable to assume that also the endpoint of the 
$CP$ lines
is first order, i.e. a triple point with two departing first
order lines, coinciding with the $T_c(\theta)$ line 
in the two adiacent branches. The picture could be modified by the presence
of dynamical fermions, because of the possible change in the
order of the transition at $\theta = 0$ and the possible appearance 
of critical endpoints in the phase diagram.
Unfortunately such scenario is not 
easily testable by lattice simulations since, contrary to what happens
for the $T$-$\theta_B$ plane, it regards a region affected by a severe
sign problem.

\section{Conclusions}
\label{conclusions}

We have presented a discussion regarding the phase diagram of pure 
gauge $SU(N)$ Yang-Mills theories in presence of a topological $\theta$ term,
based both on numerical results and on considerations related to the
large-$N$ dependence of the theory.

First, we have discussed the behavior of the deconfiment temperature
as a function of $\theta$, $T_c(\theta)$. 
The determination presented in Ref.~\cite{dene}, based on 
the method of analytic
continuation from imaginary values of $\theta$, has been improved
by performing new simulations on a finer lattice with $N_t = 10$
(which confirms the continuum extrapolation of the curvature 
at $\theta = 0$ reported in Ref.~\cite{dene}) and has been compared
with new results obtained by reweighting configurations sampled
at $\theta = 0$. As a result, we can conclude that systematic
effects related to analytic continuation and to reweighting are
under control, at least regarding the determination of the 
curvature of the critical line $R_\theta$. The final, continuum
value that we estimate for $N = 3$  is $R_\theta = 0.0178(5)$.

As a byproduct of our numerical analysis, we have explored the dependence 
of physical observables on the topological sector $Q$, showing
 that it is somewhat stronger around the transition, in particular 
for quantities directly related to deconfinement, like the Polyakov loop,
and that the 
transition temperature itself can depend on the topological background.
That may be a warning for lattice QCD studies performed in a fixed 
topological background.

Finally, in the last part of the paper, we
 have discussed the general features of the $T$-$\theta$ diagram.
Most of the discussion has been inspired by the possible analogies
and dualities (in the sense of an inversion of the low and 
high-$T$ regions) existing, also in a large $N$ perspective,
with the 
phase diagram of QCD in presence of an imaginary baryon
chemical potential. Periodicity in $\theta$ is smoothly realized in the 
high-$T$ phase and is instead associated to a multibranched structure
in the low temperature phase, where the relevant dependence is on
$\theta/N$, implying 
first order transitions which are met at odd multiples of $\theta = \pi$.
Such transitions are the analogous of the Roberge-Weiss transitions
in the high-$T$ phase of QCD in presence of an imaginary baryon chemical
potential, in both cases the change in the realization
of periodicity can be associated to a change in the relevant
degrees of freedom, carrying (topological or baryon) charge 1 or $1/N$.

\begin{acknowledgements}
We thank 
M.~Anber, F.~Bigazzi, G.~Cossu, C.~Bonati, F.~Capponi, P.~de~Forcrand, E.~S.~Fraga, 
A.~Di~Giacomo, M.~Mariti, M.~Unsal, E.~Vicari and A.~Zhitnitsky
for many useful discussions. 
We acknowledge the use of the computer facilities of the INFN Bari Computer 
Center for Science, of the INFN-Genova Section and of the CSNIV cluster 
in Pisa.
We thank the Galileo Galilei Institute for Theoretical Physics
for the hospitality offered during the workshop ''New Frontiers in
Lattice Gauge Theories".
\end{acknowledgements}


\begin{thebibliography}{9}

\bibitem{u1wit} E.~Witten, Nucl. Phys. {\bf B 156} (1979) 269.

\bibitem{u1ven} G.~Veneziano, Nucl. Phys. {\bf B 159} (1979) 213.

\bibitem{CDG-78}
C.~Callan, R.~Dashen, G.~Gross,
Phys.\ Rev.\ D\ {\bf 17}, 2717 (1978).

\bibitem{GPY-81}
D.~J.~Gross, R.~D.~Pisarski,  L.~G.~Yaffe,
Rev.\ Mod.\ Phys.\ {\bf 53}, 43 (1981).

\bibitem{schafer} 
  T.~Sch\"afer and E.~V.~Shuryak,
  Rev.\ Mod.\ Phys.\  {\bf 70}, 323 (1998)
  [hep-ph/9610451].

\bibitem{cme}
A.~Vilenkin,
  Phys.\ Rev.\ D {\bf 22}, 3080 (1980);
D.~Kharzeev, R.~D.~Pisarski and M.~H.~G.~Tytgat,
  Phys.\ Rev.\ Lett.\  {\bf 81}, 512 (1998),
  [hep-ph/9804221];
  D.~Kharzeev and A.~Zhitnitsky,
  Nucl.\ Phys.\ A {\bf 797}, 67 (2007)
  [arXiv:0706.1026 [hep-ph]];
D.~E.~Kharzeev, L.~D.~McLerran and H.~J.~Warringa,
  Nucl.\ Phys.\ A {\bf 803}, 227 (2008),
  [arXiv:0711.0950 [hep-ph]];
K.~Fukushima, D.~E.~Kharzeev and H.~J.~Warringa,
  Phys.\ Rev.\ D {\bf 78}, 074033 (2008),
  [arXiv:0808.3382 [hep-ph]].

\bibitem{azcoiti}
V.~Azcoiti, G.~Di Carlo, A.~Galante and V.~Laliena,
  Phys.\ Rev.\ Lett.\  {\bf 89}, 141601 (2002).

\bibitem{alles_1} 
  B.~Alles and A.~Papa,
  Phys.\ Rev.\ D {\bf 77}, 056008 (2008).

\bibitem{aoki_1} 
  S.~Aoki, R.~Horsley, T.~Izubuchi, Y.~Nakamura, D.~Pleiter, P.~E.~L.~Rakow, G.~Schierholz and J.~Zanotti,
  arXiv:0808.1428 [hep-lat].


\bibitem{vicari_im} 
  H.~Panagopoulos and E.~Vicari,
  JHEP {\bf 1111}, 119 (2011).

\bibitem{dene}
M.~D'Elia and F.~Negro,
  Phys.\ Rev.\ Lett.\  {\bf 109}, 072001 (2012)
  [arXiv:1205.0538 [hep-lat]].

\bibitem{theta_eff}
M.~D'Elia, M.~Mariti and F.~Negro,
  Phys.\ Rev.\ Lett.\  {\bf 110}, 082002 (2013).


\bibitem{unsal}
M.~Unsal,
  Phys.\ Rev.\ D {\bf 86}, 105012 (2012)
  [arXiv:1201.6426 [hep-th]].

\bibitem{poppitz} 
  E.~Poppitz, T.~Schäfer and M.~Ünsal,
  JHEP {\bf 1303}, 087 (2013)
  [arXiv:1212.1238 [hep-th]].

\bibitem{anber} 
  M.~M.~Anber,
  arXiv:1302.2641 [hep-th].

\bibitem{fraga}
A.~J.~Mizher and E.~S.~Fraga,
  Nucl.\ Phys.\ A {\bf 831}, 91 (2009)
  [arXiv:0810.5162 [hep-ph]].

\bibitem{kouno1}
H.~Kouno, Y.~Sakai, T.~Sasaki, K.~Kashiwa and M.~Yahiro,
  Phys.\ Rev.\ D {\bf 83}, 076009 (2011)
  [arXiv:1101.5746 [hep-ph]].

\bibitem{kouno2} 
  Y.~Sakai, H.~Kouno, T.~Sasaki and M.~Yahiro,
  Phys.\ Lett.\ B {\bf 705}, 349 (2011)
  [arXiv:1105.0413 [hep-ph]].

\bibitem{kouno3}
T.~Sasaki, J.~Takahashi, Y.~Sakai, H.~Kouno and M.~Yahiro,
  Phys.\ Rev.\ D {\bf 85}, 056009 (2012)
  [arXiv:1112.6086 [hep-ph]].


\bibitem{susc_ft} 
  B.~Alles, M.~D'Elia and A.~Di Giacomo,
  Nucl.\ Phys.\ B {\bf 494}, 281 (1997)
  [Erratum-ibid.\ B {\bf 679}, 397 (2004)];
Phys.\ Lett.\ B {\bf 412}, 119 (1997);
  Phys.\ Lett.\ B {\bf 483}, 139 (2000);
C.~Gattringer, R.~Hoffmann and S.~Schaefer,
  Phys.\ Lett.\ B {\bf 535}, 358 (2002).

\bibitem{lucini_1} 
  B.~Lucini, M.~Teper and U.~Wenger,
  Nucl.\ Phys.\ B {\bf 715}, 461 (2005).


\bibitem{vicari_ft} 
  L.~Del Debbio, H.~Panagopoulos and E.~Vicari,
  JHEP {\bf 0409}, 028 (2004).



\bibitem{brower} 
  R.~Brower, S.~Chandrasekharan, J.~W.~Negele and U.~J.~Wiese,
  Phys.\ Lett.\ B {\bf 560}, 64 (2003)
  [hep-lat/0302005].

\bibitem{aoki07} 
  S.~Aoki, H.~Fukaya, S.~Hashimoto and T.~Onogi,
  Phys.\ Rev.\ D {\bf 76}, 054508 (2007)
  [arXiv:0707.0396 [hep-lat]].

\bibitem{cossu} 
  G.~Cossu, S.~Aoki, H.~Fukaya, S.~Hashimoto, T.~Kaneko, H.~Matsufuru and J.~-I.~Noaki,
  arXiv:1304.6145 [hep-lat].


\bibitem{wit80}
E.~Witten,
  Annals Phys.\  {\bf 128}, 363 (1980).

\bibitem{ohta}
N.~Ohta,
  Prog.\ Theor.\ Phys.\  {\bf 66}, 1408 (1981)
  [Erratum-ibid.\  {\bf 67}, 993 (1982)].

\bibitem{wit98}
E.~Witten,
  Phys.\ Rev.\ Lett.\  {\bf 81}, 2862 (1998).




\bibitem{ztn1} 
  I.~E.~Halperin and A.~Zhitnitsky,
  Phys.\ Rev.\ D {\bf 58}, 054016 (1998)
  [hep-ph/9711398].

\bibitem{wit982} 
  E.~Witten,
  Adv.\ Theor.\ Math.\ Phys.\  {\bf 2}, 505 (1998)
  [hep-th/9803131].

\bibitem{saksug} 
  T.~Sakai and S.~Sugimoto,
  Prog.\ Theor.\ Phys.\  {\bf 113}, 843 (2005)
  [hep-th/0412141]; 
  Prog.\ Theor.\ Phys.\  {\bf 114}, 1083 (2005)
  [hep-th/0507073].

\bibitem{bergman} 
  O.~Bergman and G.~Lifschytz,
  JHEP {\bf 0704}, 043 (2007)
  [hep-th/0612289].


\bibitem{ztn2} 
  A.~Parnachev and A.~R.~Zhitnitsky,
  Phys.\ Rev.\ D {\bf 78}, 125002 (2008)
  [arXiv:0806.1736 [hep-ph]].

\bibitem{ztn2b} 
  A.~R.~Zhitnitsky,
  Nucl.\ Phys.\ A {\bf 813}, 279 (2008)
  [arXiv:0808.1447 [hep-ph]].

\bibitem{ztn3} 
  A.~S.~Gorsky, V.~I.~Zakharov and A.~R.~Zhitnitsky,
  Phys.\ Rev.\ D {\bf 79}, 106003 (2009)
  [arXiv:0902.1842 [hep-ph]].


\bibitem{BDPV}
C.~Bonati, M.~D'Elia, H.~Panagopoulos and E.~Vicari,
  arXiv:1301.7640 [hep-lat].


\bibitem{immu} 
M.~G.~Alford, A.~Kapustin and F.~Wilczek,
  Phys.\ Rev.\ D {\bf 59}, 054502 (1999);
A.~Hart, M.~Laine and O.~Philipsen,
  Phys.\ Lett.\ B {\bf 505}, 141 (2001);
  P.~de Forcrand and O.~Philipsen,
  Nucl.\ Phys.\ B {\bf 642}, 290 (2002);
M.~D'Elia and M.~-P.~Lombardo,
  Phys.\ Rev.\ D {\bf 67}, 014505 (2003).


\bibitem{immu_cea} 
  P.~Cea, L.~Cosmai, M.~D'Elia, C.~Manneschi and A.~Papa,
  Phys.\ Rev.\ D {\bf 80}, 034501 (2009);
  P.~Cea, L.~Cosmai, M.~D'Elia and A.~Papa,
  Phys.\ Rev.\ D {\bf 81}, 094502 (2010);
  P.~Cea, L.~Cosmai, M.~D'Elia, A.~Papa and F.~Sanfilippo,
  arXiv:1202.5700 [hep-lat].


\bibitem{vicari_b4} 
  L.~Del Debbio, H.~Panagopoulos and E.~Vicari,
  JHEP {\bf 0208}, 044 (2002).

\bibitem{nostro_b4} 
  M.~D'Elia,
  Nucl.\ Phys.\ B {\bf 661}, 139 (2003).

\bibitem{alles_2} 
  B.~Alles, M.~D'Elia and A.~Di Giacomo,
  Phys.\ Rev.\ D {\bf 71}, 034503 (2005).


\bibitem{giusti} 
  L.~Giusti, S.~Petrarca and B.~Taglienti,
  Phys.\ Rev.\ D {\bf 76}, 094510 (2007).


\bibitem{vicari_rep} 
  E.~Vicari and H.~Panagopoulos,
  Phys.\ Rept.\  {\bf 470}, 93 (2009).





\bibitem{zetaref} 
  M.~Campostrini, A.~Di Giacomo and H.~Panagopoulos,
  Phys.\ Lett.\ B {\bf 212}, 206 (1988).


\bibitem{cooling}
B.~Berg,
  Phys.\ Lett.\ B {\bf 104}, 475 (1981);
%
Y.~Iwasaki and T.~Yoshie,
  Phys.\ Lett.\ B {\bf 131}, 159 (1983);
%
S.~Itoh, Y.~Iwasaki and T.~Yoshie,
  Phys.\ Lett.\ B {\bf 147}, 141 (1984);
%
M.~Teper,
  Phys.\ Lett.\ B {\bf 162}, 357 (1985);
%
E.~-M.~Ilgenfritz {\it e al.},
  Nucl.\ Phys.\ B {\bf 268}, 693 (1986);
%
M.~Campostrini, A.~Di Giacomo, H.~Panagopoulos and E.~Vicari,
  Nucl.\ Phys.\ B {\bf 329}, 683 (1990).


\bibitem{ref:heating}
A.~Di Giacomo and E.~Vicari,
  Phys.\ Lett.\ B {\bf 275}, 429 (1992).

\bibitem{karsch_thermo}
G.~Boyd, J.~Engels, F.~Karsch, E.~Laermann, C.~Legeland, M.~Lutgemeier and B.~Petersson,
  Nucl.\ Phys.\ B {\bf 469}, 419 (1996).


\bibitem{barbour} 
  I.~M.~Barbour, S.~E.~Morrison, E.~G.~Klepfish, J.~B.~Kogut and M.~-P.~Lombardo,
  Nucl.\ Phys.\ Proc.\ Suppl.\  {\bf 60A}, 220 (1998)
  [hep-lat/9705042].

\bibitem{fodor} 
  Z.~Fodor and S.~D.~Katz,
  Phys.\ Lett.\ B {\bf 534}, 87 (2002)
  [hep-lat/0104001].


\bibitem{shsu13}
E.~Shuryak and T.~Sulejmanpasic,
  arXiv:1305.0796 [hep-ph].


\bibitem{agl03} 
  V.~Azcoiti, A.~Galante and V.~Laliena,
  Prog.\ Theor.\ Phys.\  {\bf 109}, 843 (2003)
  [hep-th/0305065].


\bibitem{rw}
A.~Roberge, N.~Weiss, Nucl. Phys. B {\bf 275}, 734 (1986).


\bibitem{sqgp}
M.~D'Elia, F.~Di Renzo, M.P.~Lombardo, Phys. Rev. D {\bf 76}, 114509 (2007).

\bibitem{rwep}
M.~D'Elia and F.~Sanfilippo,
Phys.\ Rev.\  D {\bf 80}, 111501 (2009).

\bibitem{rwep_nf3} 
P.~de Forcrand and O.~Philipsen,
Phys.\ Rev.\ Lett.\  {\bf 105}, 152001 (2010).

\bibitem{rwep2} 
C.~Bonati, G.~Cossu, M.~D'Elia and F.~Sanfilippo,
Phys.\ Rev.\ D {\bf 83}, 054505 (2011).

\bibitem{rwep3} 
C.~Bonati, P.~de Forcrand, M.~D'Elia, O.~Philipsen and F.~Sanfilippo,
arXiv:1201.2769 [hep-lat].

\bibitem{kouno}
H.~Kouno, Y.~Sakai, K.~Kashiwa and M.~Yahiro,
J.\ Phys.\ G {\bf 36}, 115010 (2009).


\bibitem{braun} 
J.~Braun, L.~M.~Haas, F.~Marhauser and J.~M.~Pawlowski,
Phys.\ Rev.\ Lett.\  {\bf 106}, 022002 (2011).

\bibitem{sakai}
Y.~Sakai, H.~Kouno, M.~Yahiro,
J.\ Phys.\ G {\bf 37}, 105007 (2010).

\bibitem{aarts}
G.~Aarts, S.~P.~Kumar and J.~Rafferty,
JHEP {\bf 1007}, 056 (2010).

\bibitem{pawlowski} 
J.~M.~Pawlowski,
AIP Conf.\ Proc.\  {\bf 1343}, 75 (2011).

\bibitem{rafferty} 
J.~Rafferty,
JHEP {\bf 1109}, 087 (2011).

\bibitem{pagura} 
V.~Pagura, D.~G.~Dumm and N.~N.~Scoccola,
Phys.\ Lett.\ B {\bf 707}, 76 (2012).

\bibitem{kashiwa} 
K.~Kashiwa, T.~Hell and W.~Weise,
Phys.\ Rev.\ D {\bf 84}, 056010 (2011).

\bibitem{morita}
K.~Morita, V.~Skokov, B.~Friman and K.~Redlich,
Phys.\ Rev.\ D {\bf 84}, 076009 (2011).

\bibitem{nagata} 
  K.~Nagata and A.~Nakamura,
  EPJ Web Conf.\  {\bf 20}, 03006 (2012)
  [arXiv:1109.0475 [hep-lat]


\bibitem{wu}
L.~-K.~Wu and X.~-F.~Meng,
  arXiv:1303.0336 [hep-lat].


\bibitem{vafawit}
C.~Vafa and E.~Witten,
  Phys.\ Rev.\ Lett.\  {\bf 53}, 535 (1984).


\bibitem{wit78} 
  E.~Witten,
  Nucl.\ Phys.\ B {\bf 149}, 285 (1979).

\bibitem{fateev} 
  V.~A.~Fateev, I.~V.~Frolov and A.~S.~Shvarts,
  Nucl.\ Phys.\ B {\bf 154}, 1 (1979).

\bibitem{berglu} 
  B.~Berg and M.~Luscher,
  Commun.\ Math.\ Phys.\  {\bf 69}, 57 (1979).

\bibitem{belavin} 
  A.~A.~Belavin, V.~A.~Fateev, A.~S.~Schwarz and Y.~S.~Tyupkin,
  Phys.\ Lett.\ B {\bf 83}, 317 (1979).


\bibitem{kraan} 
  T.~C.~Kraan and P.~van Baal,
  Phys.\ Lett.\ B {\bf 435}, 389 (1998)
  [hep-th/9806034].

\bibitem{bruck03} 
  F.~Bruckmann, D.~Nogradi and P.~van Baal,
  Nucl.\ Phys.\ B {\bf 666}, 197 (2003)
  [hep-th/0305063].

\bibitem{bruck04} 
  F.~Bruckmann, E.~M.~Ilgenfritz, B.~V.~Martemyanov and P.~van Baal,
  Phys.\ Rev.\ D {\bf 70}, 105013 (2004)
  [hep-lat/0408004].



\bibitem{diakonov} 
  D.~Diakonov and V.~Petrov,
  Phys.\ Rev.\ D {\bf 76}, 056001 (2007)
  [arXiv:0704.3181 [hep-th]]


\bibitem{unsal08} 
  M.~Unsal and L.~G.~Yaffe,
  Phys.\ Rev.\ D {\bf 78}, 065035 (2008)
  [arXiv:0803.0344 [hep-th]].

\bibitem{gorsky} 
  A.~Gorsky and V.~Zakharov,
  Phys.\ Rev.\ D {\bf 77}, 045017 (2008)
  [arXiv:0707.1284 [hep-th]].

\bibitem{dunne} 
  G.~V.~Dunne and M.~Unsal,
  Phys.\ Rev.\ D {\bf 87}, 025015 (2013)
  [arXiv:1210.3646 [hep-th]].


\bibitem{dashen} 
  R.~F.~Dashen,
  Phys.\ Rev.\ D {\bf 3}, 1879 (1971).

\bibitem{divecchia} 
  P.~Di Vecchia and G.~Veneziano,
  Nucl.\ Phys.\ B {\bf 171}, 253 (1980).

\bibitem{smilga} 
  A.~V.~Smilga,
  Phys.\ Rev.\ D {\bf 59}, 114021 (1999)
  [hep-ph/9805214].

\bibitem{tytgat} 
  M.~H.~G.~Tytgat,
  Phys.\ Rev.\ D {\bf 61}, 114009 (2000)
  [hep-ph/9909532].

\bibitem{creutz} 
  M.~Creutz,
  Phys.\ Rev.\ Lett.\  {\bf 92}, 201601 (2004)
  [hep-lat/0312018]; 
  arXiv:1306.1245 [hep-lat].

\bibitem{metlit}
M.~A.~Metlitski and A.~R.~Zhitnitsky,
  Phys.\ Lett.\ B {\bf 633}, 721 (2006)
  [hep-ph/0510162]; Nucl.\ Phys.\ B {\bf 731}, 309 (2005)
  [hep-ph/0508004].

\bibitem{boomsma1}
D.~Boer and J.~K.~Boomsma,
  Phys.\ Rev.\ D {\bf 78}, 054027 (2008)
  [arXiv:0806.1669 [hep-ph]].


\bibitem{boomsma2}
J.~K.~Boomsma and D.~Boer,
  Phys.\ Rev.\ D {\bf 80}, 034019 (2009)
  [arXiv:0905.4660 [hep-ph]].



\end{thebibliography}
\end{document}